\documentclass[aps,prl,twocolumn]{revtex4-1}
\usepackage{graphicx}
\usepackage{float}
\usepackage{times}
\usepackage{color}
\input pdfcolor.tex
\usepackage{graphicx}
\usepackage{epsfig,color}

\begin{document}

\author{Sauri Bhattacharyya$^{1}$, Sankha Subhra Bakshi$^{1}$, 
	Saurabh Pradhan$^{2}$ and Pinaki Majumdar$^{1}$}

\affiliation{
$^1$~Harish-Chandra Research Institute, HBNI,
Chhatnag Road, Jhunsi, Allahabad 211 019, India\\
$^2$~Department of Physics and Astronomy, Uppsala University,
751 05 Uppsala, Sweden }

\title{
Strongly anharmonic collective modes in a coupled 
electron-phonon-spin problem}

\date{\today}

\begin{abstract}
We solve for the finite temperature collective mode dynamics in the 
Holstein-double exchange problem, using coupled Langevin equations for 
the phonon and spin variables. We present results in a strongly anharmonic 
regime, close to a polaronic instability. For our parameter choice the system 
transits from an `undistorted' ferromagnetic metal at low temperature to a 
structurally distorted paramagnetic insulator at high temperature, through 
a short range charge ordered (CO) phase near the ferromagnetic crossover
at $T_{FM}$. The  small amplitude harmonic phonons at low temperature cross 
over to large amplitude dynamics around $0.5 T_{FM}$ due to thermally 
generated short range correlated polarons. The rare thermal ``tunneling'' 
of CO domains generates a hitherto unknown momentum selective spectral weight 
at very low energy.  We compare our results to inelastic neutron data in the 
manganites and suggest how the singular low energy features can be probed.
\end{abstract}


\maketitle

Collective modes play a crucial role in dictating low-energy
spectral and transport properties in correlated electron systems
\cite{dag,man,cup}.
The most detailed information about them comes
from inelastic neutron scattering (INS) 
experiments \cite{ins1,ins2,
weber1,weber2,weber3,weber4,kajimoto,dean,
dai,zhang,chatterji,ye,moussa,ulbrich,helton},
which probe the 
momentum resolved spectrum of lattice, magnetic, or density
fluctuations. The interpretation of INS results has depended,
traditionally, on schemes like the random phase approximation
(RPA), for phonons, or the $1/S$ expansion for spins. These
methods 
are meant to access low amplitude fluctuations and are limited 
to low temperature.

Correlated electron systems often show thermal phase transitions, 
or strong short range correlated distortions
\cite{dag,man,cup}, 
where the low temperature `linearised' dynamics is no longer useful 
in describing the modes.  While methods like 
time dependent Ginzburg-Landau (TDGL) theory \cite{TDGL1,TDGL2} 
or full scale `molecular dynamics' (MD) \cite{MD1,MD2} are available 
in classical systems, equivalent real time methods 
are rare in quantum systems. 
To address collective mode physics in 
these systems a method should handle the
strong interactions reliably, capture large
dynamical fluctuations in real time, be sensitive to 
spatial correlations, and access finite temperature. 


In this paper we demonstrate a `real space', real time
method that probes the dynamics of phonons and spins 
strongly coupled to an electron system,
handles strong interactions non perturbatively,
and accesses thermal physics.
We deliberately choose an `operating point' where 
anharmonic effects in the phonons 
- arising from mode coupling, polaron `tunneling', 
and magnetic fluctuations - 
are large.  This allows us to demonstrate 
the uniqueness of the method, and also address 
dynamics in the manganites as a non trivial test case.

The manganites provide a concrete template for
thermally induced anharmonic fluctuations.
The itinerant $e_g$ electrons in these materials are
strongly coupled to lattice modes via Jahn-Teller (JT) coupling 
\cite{man}, and to $t_{2g}$ based core spins via 
large Hund's coupling \cite{man,millis}.
The JT coupling favours polaron formation (and electron
localisation), while the Hund's coupling favours a 
ferromagnetic metallic (FM-M) state.
There are materials, {\it e.g}, La$_{1-x}$Ca$_x$MnO$_3$,
with $x \sim 0.3$, where thermally induced polaronic
distortions convert 
a homogeneous low $T$ FM-M to a structurally
distorted polaronic insulator above magnetic
transition at $T_{FM}$. 
The structural and transport features of this problem
\cite{man,uru1,uru2,tok-bicr,cluster-cmr-expt,liu,coey,saitoh} 
has seen much analysis 
\cite{dag-dis,furu-dis,kumar} over the last two decades
while the thermal dynamics remained virtually unaddressed.

Experiments exist on the phonon 
\cite{weber1,weber2,weber3,weber4}
and spin dynamics
\cite{dai,zhang,chatterji,ye,moussa,ulbrich,helton}.
INS results reveal that (i)~in 
La$_{1-x}$Sr$_x$MnO$_3$ with $x\sim 0.2-0.3$,
the transverse acoustic phonons show anomalous softening and 
broadening \cite{weber1,weber2} on heating through $T_{FM}$,
while (ii)~the magnons along (100) and (110) directions 
in La$_{0.7}$Ca$_{0.3}$MnO$_{3}$ \cite{dai,ye,helton} and 
La$_{0.8}$Sr$_{0.2}$MnO$_{3}$ \cite{zhang,moussa} 
show large linewidths and softening near $T_{FM}$, which can't 
be explained using a simple spin model.

Our work reveals that for a model that involves thermally 
induced metal-insulator transition via polaron formation,
the dynamical features above arise naturally,
and, additionally, are accompanied by rare events relating
to polaron tunneling. This  leads to a remarkable low energy
signature in the spectrum. 
We use a Langevin dynamics approach 
\cite{chern,sauri}, new to these problems,
a Holstein model, rather than Jahn-Teller, for the
phonons, a large Hund's coupling to drive ferromagnetism,
and solve a two dimensional 
\cite{crossover} electron problem.
The electron-phonon (EP) coupling is chosen 
so that the $T=0$ system is just below the polaronic 
instability. The bare phonon frequency is $\Omega$. 
Our main results are the following.

(A).~{\it Phonons:}
The phonon spectrum exhibits the expected `RPA' dispersion
at $T=0$, with resolution limited lineshapes, but increasing $T$
leads to three prominent effects: 
(a)~The growing phonon distortions, and the associated electron
density, order in a short range pattern with 
wavevector ${\bf Q} \sim (\pi,\pi)$ and the phonon dispersion
for momentum ${\bf q} \sim {\bf Q}$ softens significantly.
(b)~The damping $\Gamma_{\bf q}$ is strongest for ${\bf q}
\sim {\bf Q}$ and grows rapidly with $T$ due to a combination
of anharmonic phonon interaction and magnon-phonon coupling.
(c)~The short range correlated structures have their own 
slow dynamics - involving thermally assisted tunneling - 
and this generates visible spectral weight at frequency
$\omega \ll \Omega$~for~${\bf q} \sim {\bf Q}$.

\begin{figure*}[t]
\centerline{~~~~~~~~~~
\includegraphics[width=13.0cm,height=2.6cm]{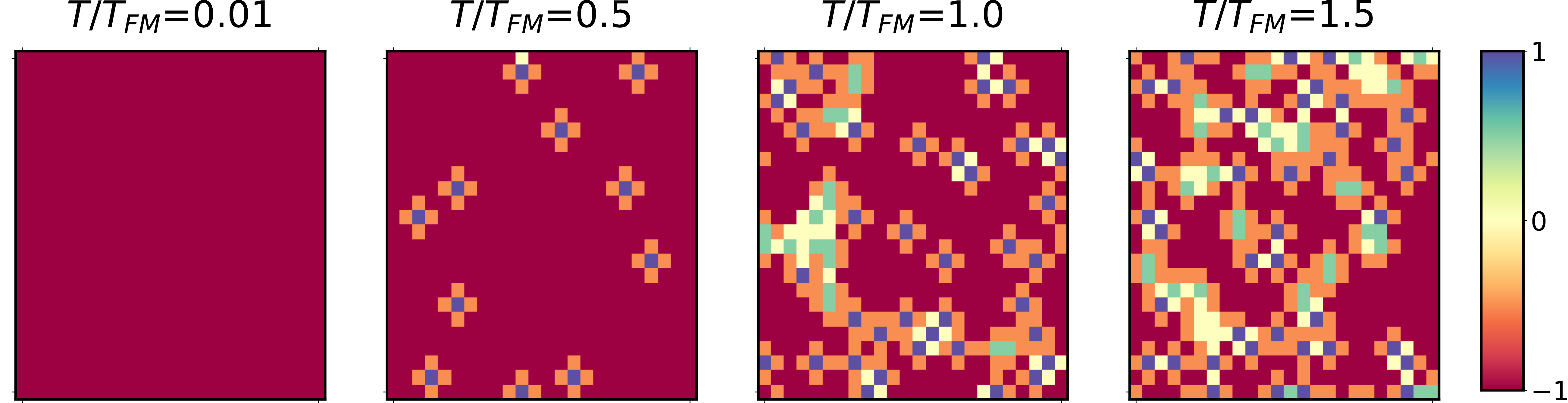} }~~~~~~
\vspace{.1cm}
\centerline{
\includegraphics[width=13.0cm,height=2.4cm]{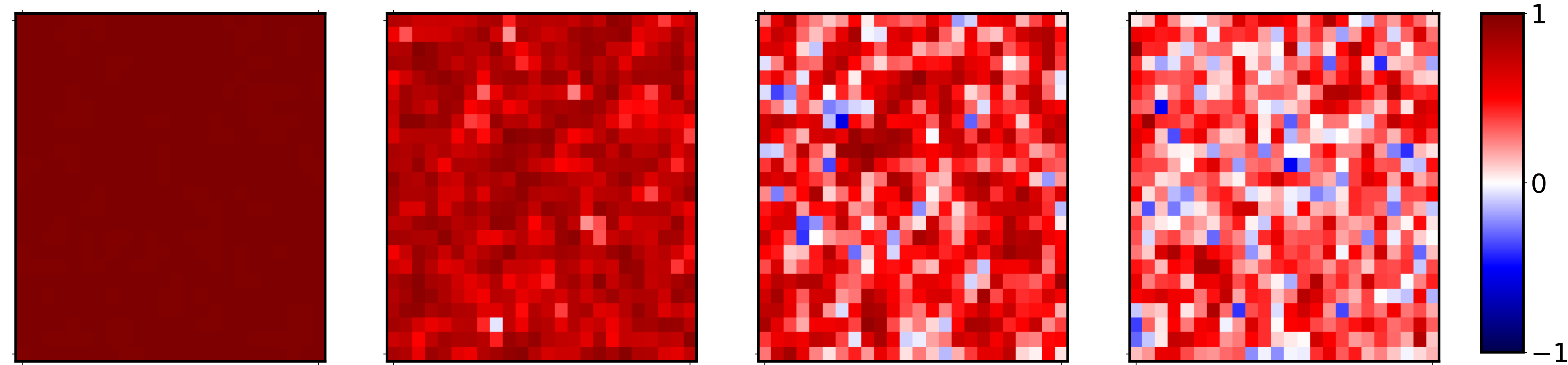} }
\centerline{
~~~~~
\includegraphics[width=13.9cm,height=2.7cm]{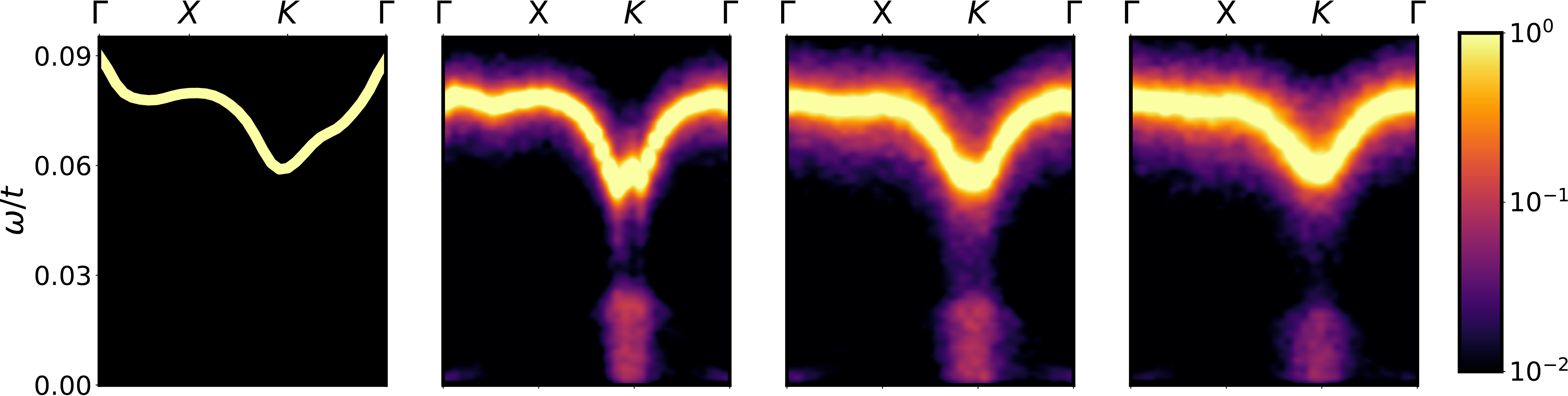} }
\centerline{
~~~~~
\includegraphics[width=13.5cm,height=2.7cm]{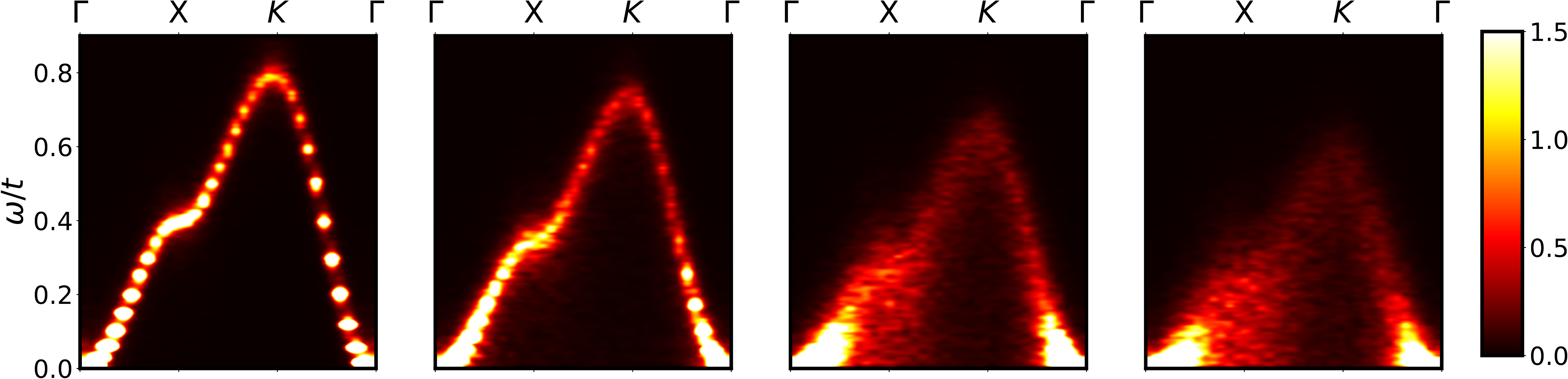} 
}
\caption{Phonon and spin snapshots at temperatures $T/T_{FM}
= 0.01,~0.5,~1.0,~1.5$ and the corresponding spectral maps.
The $T$ are chosen to represent low, intermediate, `critical',
and high temperature regimes.  First row: snapshot of the phonon
field $x_i(t)$, showing
the change from an undistorted low $T$ state to a progressively
large distortion checkerboard correlated state with temperature.
Second row: snapshot of nearest-neighbour spin correlation:
$O_i(t)= {1 \over 4} \sum_{\delta} {\vec S_i(t)}. 
{\vec S_{i + \delta}(t)}$,
showing the evolution from a perfect FM to a disordered
state on heating through $T_{FM}$.  Third row: phonon power
spectrum ${1 \over T} 
\vert X({\bf q},\omega)\vert^2$ for ${\bf q}$ varying
along $\Gamma-X-K-\Gamma$ in the Brillouin zone.  We see a thermally
induced softening and broadening for modes with ${\bf q} \sim (\pi,\pi)$,
alongwith an unexpected $\omega \rightarrow 0$ feature for $T \gtrsim
0.5T_{FM}$.  Fourth row: magnon power spectrum ${1 \over T} 
\vert S({\bf q},
\omega)\vert^2$, starting from a Heisenberg-like spectrum with $J_{eff}
\sim0.1t$, shows overall softening and large dampings near
$(\pi,0)$ and $(\pi,\pi)$ on heating up.  }
\end{figure*}

(B).~{\it Magnons:}
The low temperature magnons are as expected in a ferromagnet,
with $\omega_s({\bf q}) \sim J_{eff}(2 - \cos(q_x a) - \cos(q_y a))$,
where $J_{eff} \sim 0.1t$, 
and resolution limited widths. 
Temperature brings in two features: (a)~the spectrum narrows
and the mean dispersion shows a softening that is roughly
linear in $T$, and (b)~The damping stays small till $T
\sim 0.5 T_{FM}$ and then shows a dramatic increase.
Unlike phonons, for whom the principal weight remains
at $\omega \sim \Omega$, the high temperature magnon 
lineshape is very broad.

{\bf Model and method:}
We study the Holstein-double exchange (HDE)
model on a two-dimensional 
square lattice. 
\begin{eqnarray}
H &=& -t\sum_{<ij>}^{\sigma} c^{\dagger}_{i\sigma}c_{j\sigma} 
-J_{H}\sum_{i}\vec{S}_{i}.\vec{\sigma}_{i}  
- \mu\sum_{i}n_{i} \cr
 && ~~~~-g\sum_{i}n_{i}x_{i} 
    +\sum_{i}(\frac{p^2_{i}}{2M} + \frac{1}{2}Kx^2_{i})
\end{eqnarray}
We study a nearest neighbour model 
with $t=1$ at density $n=0.40$.
$K$ and $M$ are the local 
stiffness and mass, 
respectively, of the optical phonons, 
and 
$g = 1.40$
is the electron-phonon coupling. 
We set $K=1$. In this paper, we 
report studies for $\Omega = \sqrt{K/M} = 0.1t$, 
which is a reasonable value for real materials. 
$\vec{S}_{i}$'s are `core spins', assumed
to be large and classical.
The chemical potential $\mu$ is varied to maintain 
the electron density at the required value. 
We work in the Hund's coupling limit 
$J_{H}/t \gg 1$.

The thermal dynamics of the phonons and spins is solved using
the coupled Langevin equations (see Supplement):
\begin{eqnarray}
M{ {d^2x_i} \over dt} & = & 
-D_{ph} { {dx_i} \over dt}   - K x_i - 
{ {\partial {\langle H_{el} \rangle }}   \over {\partial x_i }} 
+ \xi_i(t) \cr \cr
{{d {\vec S}_i} \over {dt} }~~ & =&
-\vec{S}_{i}\times({{\partial
{\langle H_{el} \rangle }}\over {\partial \vec{S}_i }}
+ \vec{h}_{i}) + ~D_{s}\vec{S}_{i}\times (\vec{S}_{i}
\times{{\partial {\langle H_{el} \rangle }}\over
{\partial \vec{S}_{i} }} ) \cr
\cr
H_{el}~ & = & \sum_{ij}(t_{ij} - \mu \delta_{ij}) 
\gamma^{\dagger}_{i}\gamma_{j} - g\sum_in_ix_i
\cr
t_{ij}/t ~&= & \sqrt{(1+\vec{S}_i.\vec{S}_j)/2}
\end{eqnarray}
The phonon equation \cite{sauri}
involves inertia, damping, an effective 
force from the electronic energy, and noise.
The noise satisfies the fluctuation-dissipation theorem (FDT)
and is specified by
$ \langle \xi_{i}(t)\rangle = 0,~~
\langle \xi_{i}(t)\xi_{j}(t^{\prime}) \rangle = 2D_{ph} 
k_{B}T\delta_{ij}\delta(t-t^{\prime})$.
The spin dynamics follows a Landau-Lifshitz-Gilbert-Brown
(LLGB) equation \cite{brown}.   
The first term on the right hand side of the
spin equation is the torque 
and the second term is the Gilbert damping.
The noise ${\vec h}_i$ also 
satisfies FDT but enters in a `multiplicative' form,
crossed with the spin field ${\vec S}_{i}$ itself. 
The spin evolution conserves $\vert {\vec S}_i \vert$.

There are multiple timescales involved. We set the bare
oscillation period for the phonons,
$\tau_{ph}=2\pi/\Omega$, as the unit of time.
The low $T$ relaxation time for phonons is 
$2M/D_{ph} \sim60\tau_{ph}$, for $D_{ph}=0.05t$. 
For magnons, the 
typical time period is set by $\tau_{s}=
1/J_{eff}$ of the effective Heisenberg model, roughly $10/t$. 
The low $T$ magnon relaxation timescale is 
$D_{s}^{-1} \sim 4\tau_{s}$. 
For our parameter choice  $\tau_{ph} \sim 6\tau_{s}$.
The largest timescale is for phonon relaxation,
at $60 \tau_{ph}$, and the smallest is for magnetic
oscillations at $\sim (1/6) \tau_{ph}$. The numerical
scheme has to use time discretisation and overall 
runtime keeping these in mind.
The choice of $D_{ph}$ and $D_s$ is discussed in the 
Supplement.

The evolution equations are numerically integrated using the 
Euler-Maruyama \cite{kloeden}
scheme for phonons and a Suzuki-Trotter 
decomposition based method \cite{ma} for spins. The time
step is $\Delta t=1.6 \times 10^{-4}\tau_{ph}$. We typically 
ran our simulations on systems of size 
$24\times24$ for $10^{7}$ steps, {\it i.e} $\sim 10^3 \tau_{ph}$. 
This ensured that we had enough time for equilibration and  
enough frequency resolution to compute the power spectrum. 

{\bf Results:}
We organize the results in three parts: 
(a)~$T$ dependence of the typical instantaneous 
phonon  and spin backgrounds and gross spectral features, 
(b)~the phonon and magnon lineshape at a few momenta, and 
(c)~comparison of our results with inelastic neutron data
on the manganites.

Fig.1 correlates the typical phonon and spin backgrounds, in the
upper two rows,  obtained as instantaneous Langevin configurations,
with the momentum resolved power spectrum of the phonons and spins
in the bottom two rows. 
The top row shows phonon configurations $\{x_i\}$.
We see an undistorted state at low $T$ (left
panel) gradually forming patches 
of checkerboard ordered large 
distortions on heating up.
These patches proliferate in the critical regime.
The changing $x_i$ background 
leads to a rise in the density correlation function 
$S_n({\bf q})$ at ${\bf q} =  (\pi,\pi)$,
shown in the Supplement. The second row shows 
snapshots of 
the nearest neighbour summed overlap
$O_i(t) = {1 \over 4} \sum_{\delta} {\vec S_i(t)}. {\vec S_{i + \delta}}(t)$, 
which indicate a ferromagnetic low $T$ 
state, and progressively spin disordered configurations 
on heating across $T_{FM}$. The $T$ dependence of the 
associated FM peak, $S_{s}(0,0)$, in the structure factor,  
is shown in the Supplement. The magnetic disorder aids lattice 
polaron formation by suppressing the hopping.

The third row shows ${1 \over T} 
\vert X({\bf q},\omega) \vert^2$,
where $X({\bf q},\omega) = \sum_i e^{i {\bf q}.{\bf r}_i} 
\int dt e^{-i \omega t} x_i(t) $.   
Sample behaviour of $x_i(t)$ in different $T$ regimes
is shown in the Supplement.
The low temperature phonon spectrum (first column)
is accessible through a harmonic theory, equivalent to RPA in
the quantum context, with intersite phonon
correlations arising via the `bare' 
electronic polarisability $\Pi_{0}({\bf q})$. 
The phonon dispersion has a form $\omega^0_{ph}({\bf q}) 
\sim \sqrt{(K+ g^2\Pi_0({\bf q}))/M}$
and the damping $\Gamma_{ph}({\bf q})$ 
of these `normal modes' is $\sim D_{ph}/M$.
On heating up, the typical $x_i$ increase in magnitude and
the spectrum 
displays three distinct features- 
(i)~increased damping due 
to anharmonicity induced coupling between normal modes,
(ii)~`softening' of the dispersion near ${\bf q}= (\pi, \pi)$,
related to enhanced CO correlations, 
and (iii)~the appearance of
spectral weight at low frequencies,  $\omega \ll \Omega$! 
The low-energy feature arise from rare tunneling of 
checkerboard correlated
patches that lead to large local `switching' of the $x_i$.
The low energy weight reduces when $T \gg T_{FM}$ where large amplitude
oscillations and tunneling events can no longer be distinguished.
We will discuss the impact of the magnetic degrees of freedom on
the phonons later in the paper.

\begin{figure}[b]
\centerline{
\includegraphics[width=4.0cm,height=3.4cm]{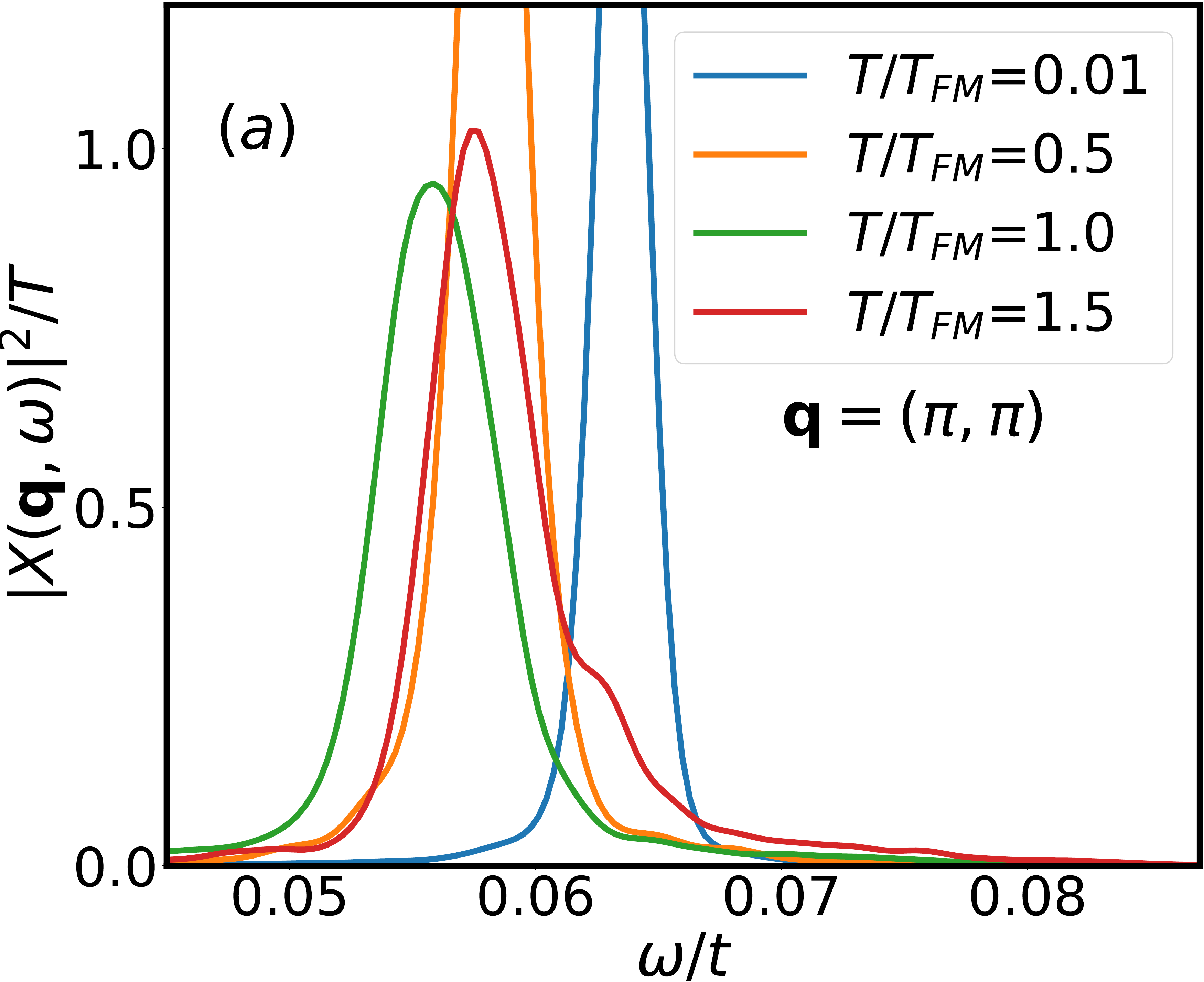}
\hspace{.2cm}
\includegraphics[width=4.0cm,height=3.4cm]{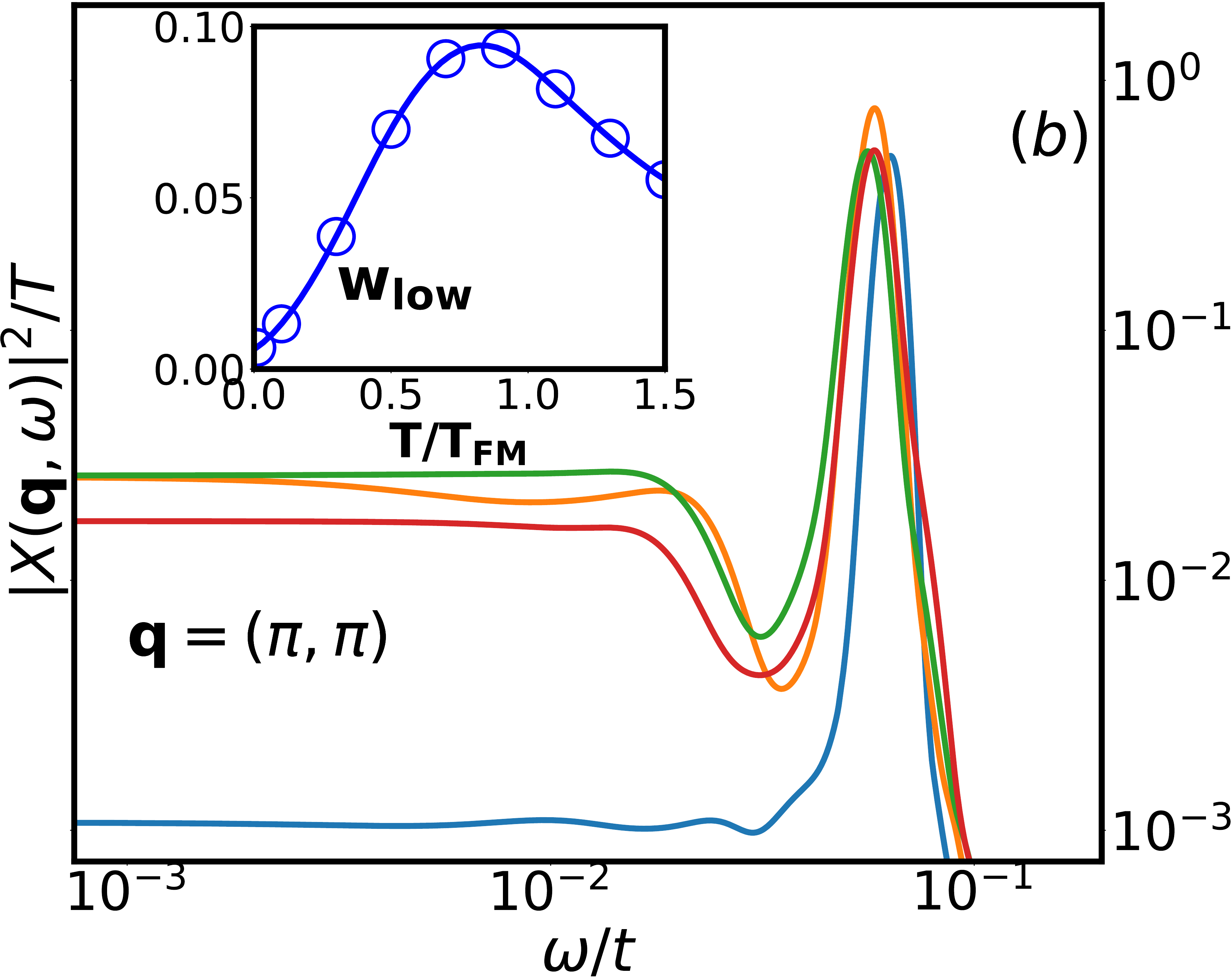}
}
\centerline{
\includegraphics[width=4.0cm,height=3.4cm]{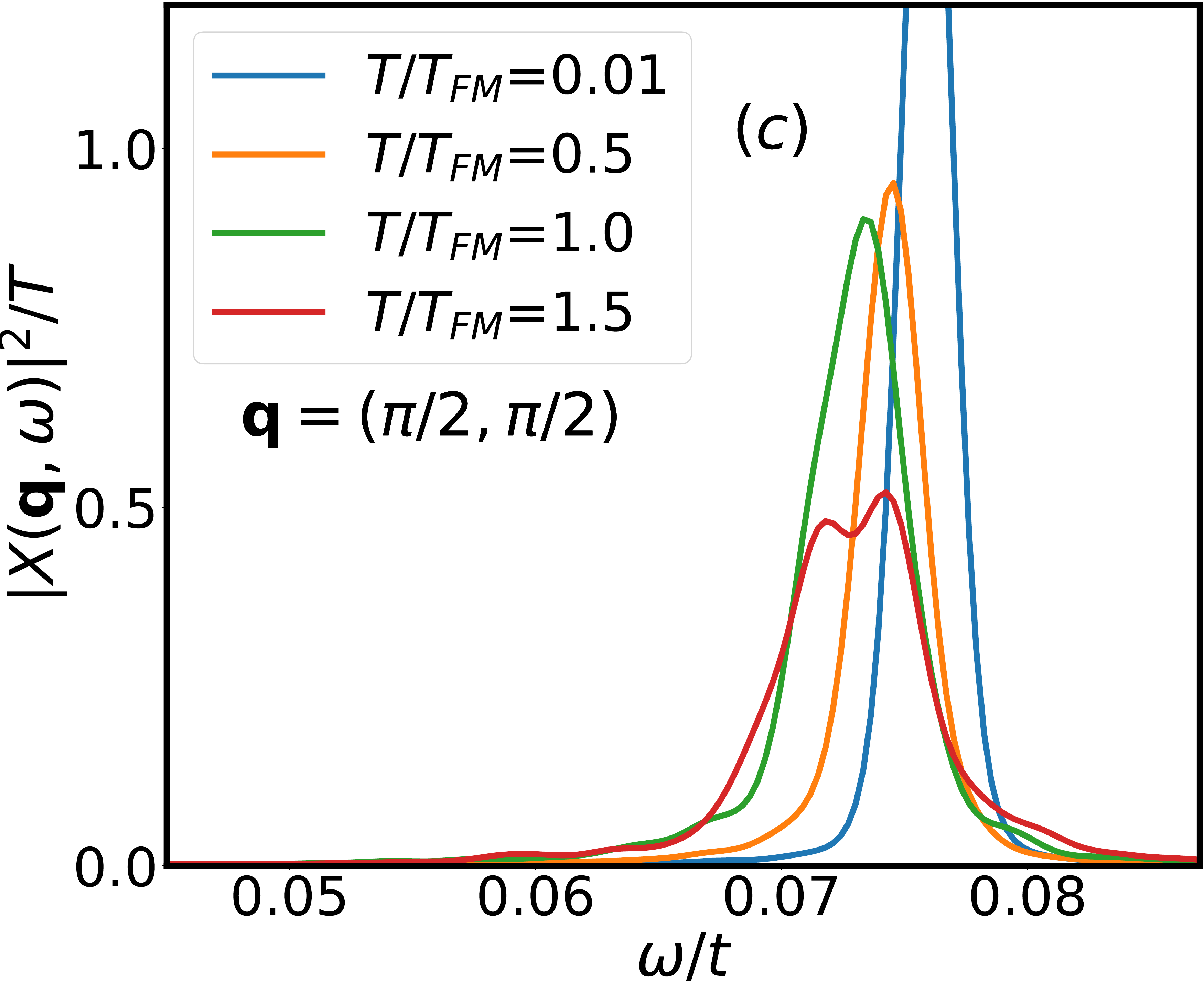}
\hspace{.2cm}
\includegraphics[width=3.9cm,height=3.4cm]{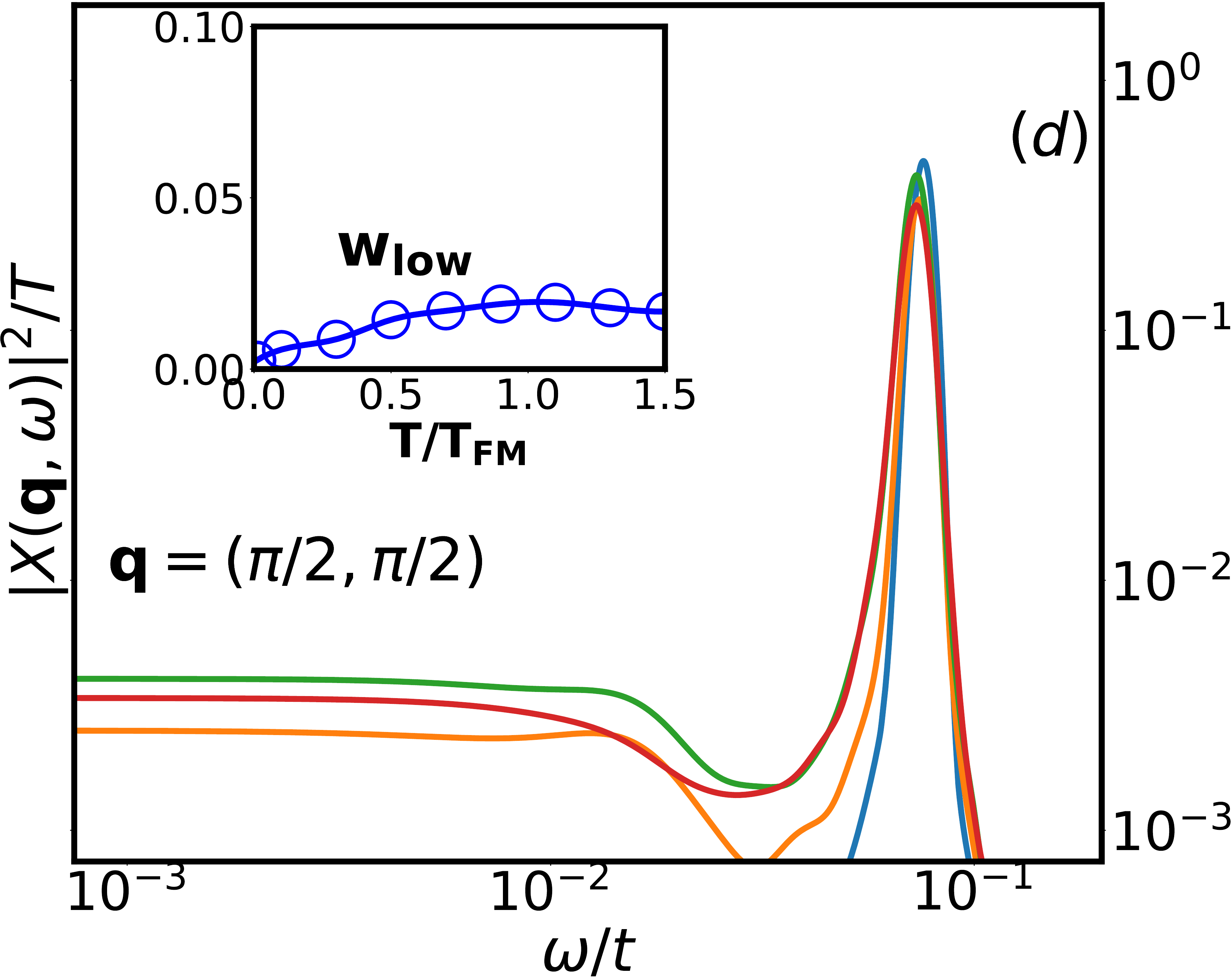}
}
\centerline{~~~~~
\includegraphics[width=4.1cm,height=3.4cm]{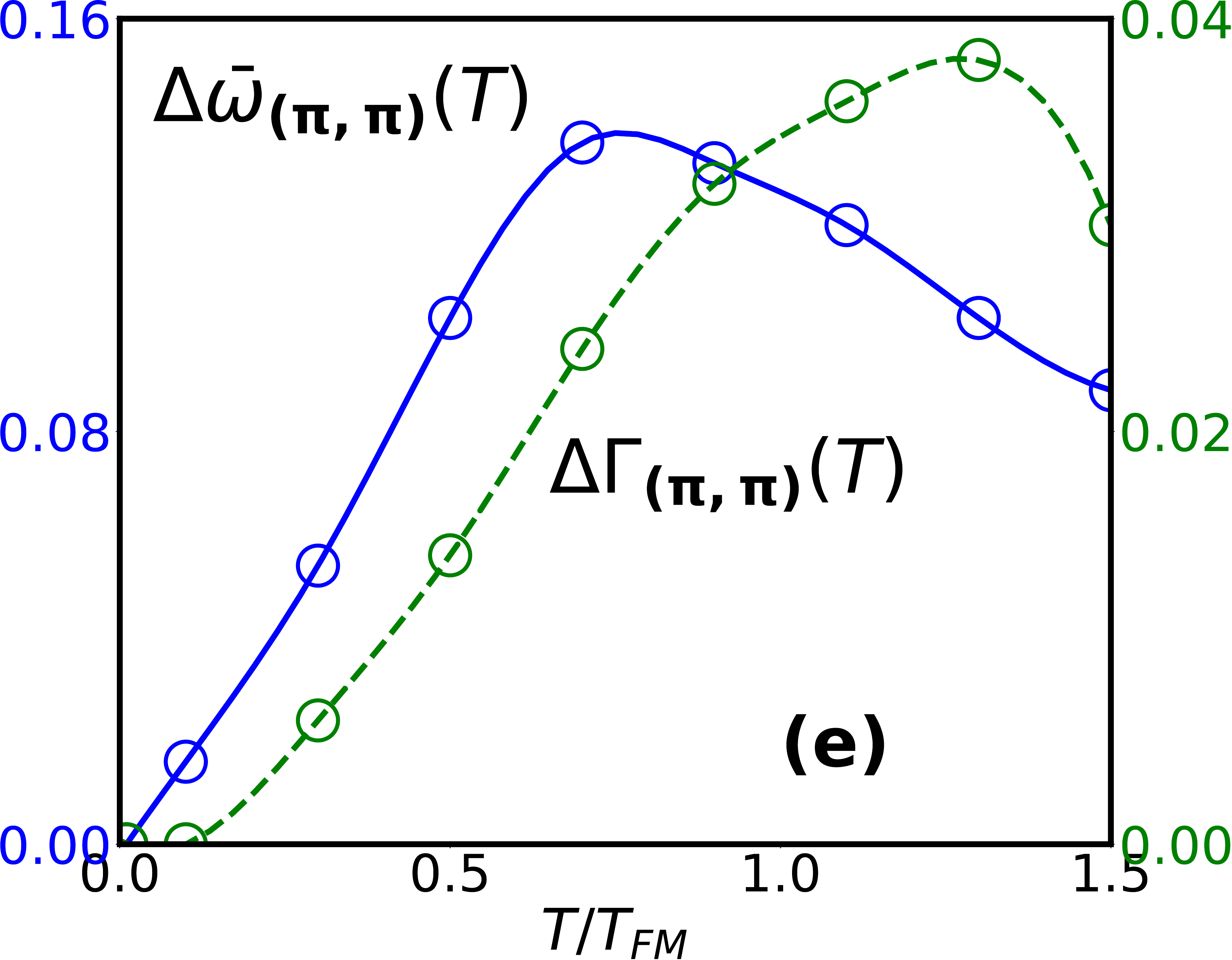}
\hspace{-.1cm}
\includegraphics[width=3.9cm,height=3.4cm]{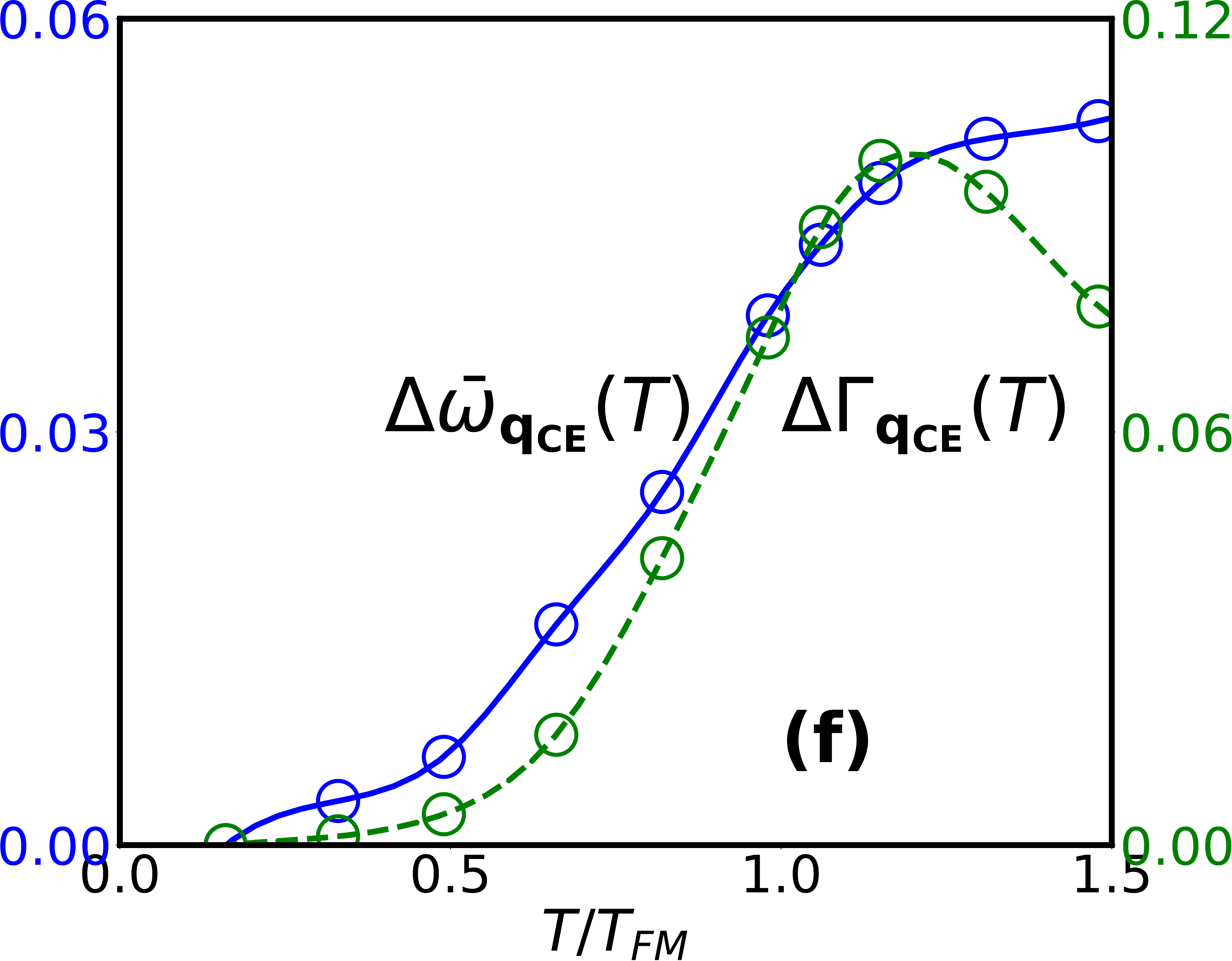}
~
}
\caption{Top row: Phonon lineshape for ${\bf q}=(\pi,\pi)$
on a linear (a) and logarithmic (b) scale.
The temperatures chosen are- $T/T_{FM}={0.01,0.5,1.0,1.5}$.
We observe a $\sim 15\%$ softening of mode frequency and a sharp
increase in the linewidth with $T$ in (a). The accumulation of
low-energy weight is emphasized in (b), where the inset
shows a detailed $T$ dependence. Middle: the same analysis is
repeated for ${\bf q}=(\pi/2,\pi/2)$ in (c) and (d).  Similar
trends persist with much reduced extent.  Bottom: Theoretically
extracted `softening' $\Delta\bar{\omega}_{(\pi,\pi)}(T) =
\bar{\omega}_{(\pi,\pi)}(0) - \bar{\omega}_{(\pi,\pi)}(T)$ and thermal
broadening $\Delta\Gamma_{(\pi,\pi)}(T) = \Gamma_{(\pi,\pi)}(T) -
\Gamma_{(\pi,\pi)}(0)$ in (e) is compared to corresponding
quantities for ${\bf q} ={\bf q}_{CE}$ from experiments (f).
Qualitative trends are similar.
}
\end{figure}

The fourth row shows ${1 \over T} 
\vert \vec{S}({\bf q},\omega) \vert^2$, 
where
$\vec{S}({\bf q},\omega) = \sum_i e^{i {\bf q}.{\bf r}_i} 
\int dt e^{-i \omega t} \vec{S}_i(t) $. 
%
%
The low temperature spectrum corresponds to 
FM spin waves, with $\omega^0_s({\bf q}) 
\sim J_{eff}(2 - \cos(q_x a) - \cos(q_y a))$ 
and can be reproduced
using a nearest neighbour 
FM Heisenberg model with $J_{eff} \approx 0.1t$.
Increasing $T$ reveals a suppression of the intensity 
and a slow increase in magnon damping. 
Beyond  $\sim 0.5 T_{FM}$ the magnon lines 
broaden rapidly, 
notably near $(\pi,0)$ and $(\pi,\pi)$. Beyond $T_{FM}$, most
of the modes are diffusive in nature, except near the zone center.
These thermal trends are qualitatively similar to the 
Heisenberg model \cite{vineyard,takahashi} and also to a pure double 
exchange model at the same density. We will discuss the 
relative insensitivity of magnons to phonon physics later.

Fig.2 examines phonon lineshapes in detail at two momenta, 
${\bf q} = (\pi,\pi)$ and $(\pi/2,\pi/2)$.
Panel 2(a) focuses on the `high energy' part of the spectrum,
$\omega \sim \Omega$, at ${\bf q} = (\pi,\pi)$, while 2(b) uses
a logarithmic frequency and amplitude scale to show the full
${\bf q} = (\pi,\pi)$ data. 
In 2(a), we see a striking enhancement of broadening 
in the high-energy part of the spectrum on heating from the
$T = 0.01T_{FM}$ to $T \sim T_{FM}$.
A reduction of the mean frequency is also observed. 
To clarify the behaviour at $\omega \ll \Omega$
panel 2(b) depicts the full spectrum in a log-log plot. This 
reveals 
the low frequency spectral weight arising from polaron tunneling.
The inset shows the $T$ dependence of the low frequency weight
$w_{low}({\bf q}, T) = \int_0^{\Omega_{low}} d\omega
\vert X({\bf q}, \omega) \vert^2$, normalised by the the full
weight. We use $\Omega_{low} = 0.4\Omega$.
The weight $w_{low}$ is non monotonic in $T$ with a reasonable
maximum value $\sim 10\%$.
2(c)-(d) repeats  the same analysis for $(\pi/2,\pi/2)$, 
which is considerably separated from the `CO'  wavevector.
The trends are similar to  ${\bf q} = (\pi,\pi)$ 
but the peak $w_{low}$ is much smaller, $\sim 3\%$.

Figs.2(e) and 2(f) show a comparison of the 
softening and damping inferred from our $(\pi,\pi)$ phonon lineshape 
with that extracted from experimental data at the `CE ordering' wavevector
${\bf q} ={\bf q}_{CE}$. 
The experimental data is on {\it acoustic} phonons,
but its has been argued that the behaviour should be similar to that
of the JT phonons. Both theory and  
experimental results are normalised by the respective low $T$ bandwidth,
and temperatures are scaled by the respective $T_{FM}$,
$\sim 0.1t$ in the model and $305$K in experiments. 
The $T$ dependence in panels (e) and (f) share similarities
but actual numbers differ 
\cite{expt-theory} 
by $\sim $ a factor of 2. We
discuss the comparison in more detail later.
To the extent we know, experiments
have not probed the low frequency part of the spectrum.

Figs.3 is focused on magnons, at ${\bf q} = (\pi, 0)$ and $(\pi, \pi)$.
The lineshapes show that the sharp dispersive feature 
for $T \lesssim 0.5T_{FM}$ and then rapid broadening 
as $T \rightarrow T_{FM}$. 
The $(\pi,0)$ mode softens much less than the mode at $(\pi, \pi)$.
In 3(c) and 3(d), the detailed temperature dependence of 
softening, $\Delta\omega_{\bf q}(T)$, and damping, $\Delta\Gamma_{\bf q}(T)$,
are shown.  The normalizing energy scale is the low $T$ 
magnon bandwidth ($\sim0.8t$ in our case). 
We have not been able to find systematic temperature dependence data
on magnon lineshapes in the manganites, although a body of
results \cite{dai,zhang,ye,helton}
point out low temperature magnon `anomalies' in these materials.

{\bf Discussion:}
Having presented the results, in
what follows we provide an analysis of the
features in Figs.1-3, and point out 
where our results match with, differ from, and go 
beyond measurements  in the manganites.

We broadly observe four phonon regimes- 
(i)~harmonic, $\sim0-0.1T_{FM}$, 
(ii)~anharmonic, $\sim0.1T_{FM} - 0.3T_{FM}$, 
(iii)~polaronic, $\sim0.3T_{FM}-1.5T_{FM}$,
and (iv)~large oscillations, $\gtrsim 1.5 T_{FM}$, 
in terms of real-time dynamics 
(see Supplement for $x_i(t)$ data).
The harmonic to anharmonic crossover is reflected in the $T$
dependence of $\Gamma({\bf q})$ due to mode coupling. In
the polaronic regime the distortions increase and we
observe `burst like' events - anticorrelated between
nearest neighbour sites. For $T \gtrsim 1.5T_{FM}$ the
oscillations are even larger but the 
spatial correlations begin to weaken.
Regimes (iii) and (iv) contain appreciable effect of magnetic
disorder, which results in a peak in
$\Gamma_{ph}({\bf q},T)$ for $T\sim T_{FM}$ for the present case 
- in contrast to a pure Holstein model. 


The observed magnon spectra are similar to those of a 
nearest-neighbour Heisenberg model. The `square root' 
renormalization of stiffness, and phonon couplings, are 
both seemingly irrelevant.
There are broadly three magnon regimes- 
(i)~free,  $\sim0-0.5T_{FM}$, (ii)~interacting, $\sim0.5T_{FM}-T_{FM}$,
and (iii)~diffusive, $\gtrsim T_{FM}$.
The ${(1 + \langle S_i.S_j \rangle )}^{1/2}$ 
factor varies by $\sim15\%$ from $0-T_{FM}$, while 
$\langle\gamma_{i}^{\dagger}\gamma_{j}\rangle$
is almost $T$ independent. Due to this an essentially
phonon insensitive Heisenberg description arises.
The overall picture holds even in 
presence of a {\it small} $J_{AF}$, whose main effect is  
bandwidth reduction at low $T$.

\begin{figure}[t]
\centerline{
\includegraphics[width=4.0cm,height=3.3cm]{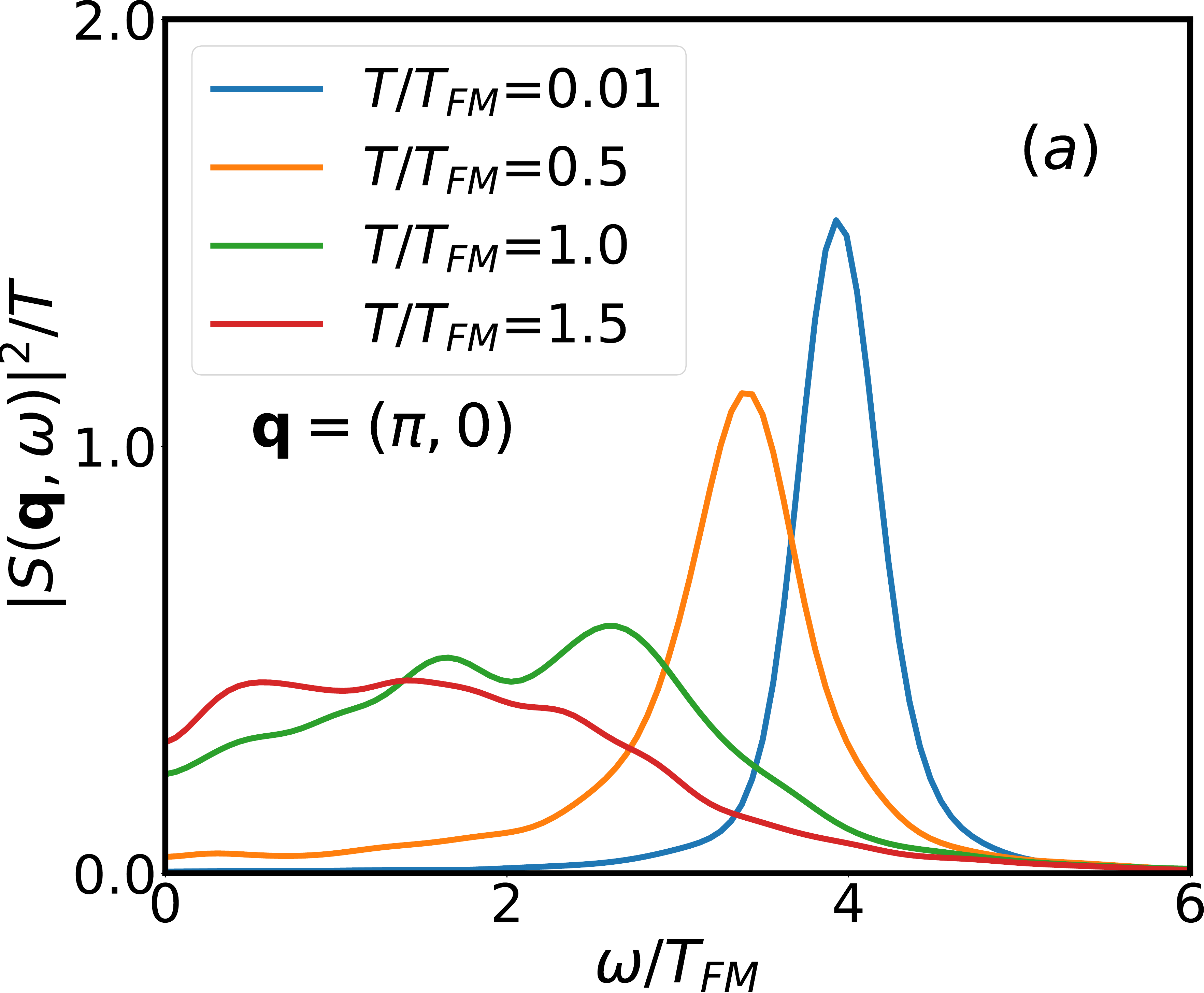}
\includegraphics[width=4.0cm,height=3.3cm]{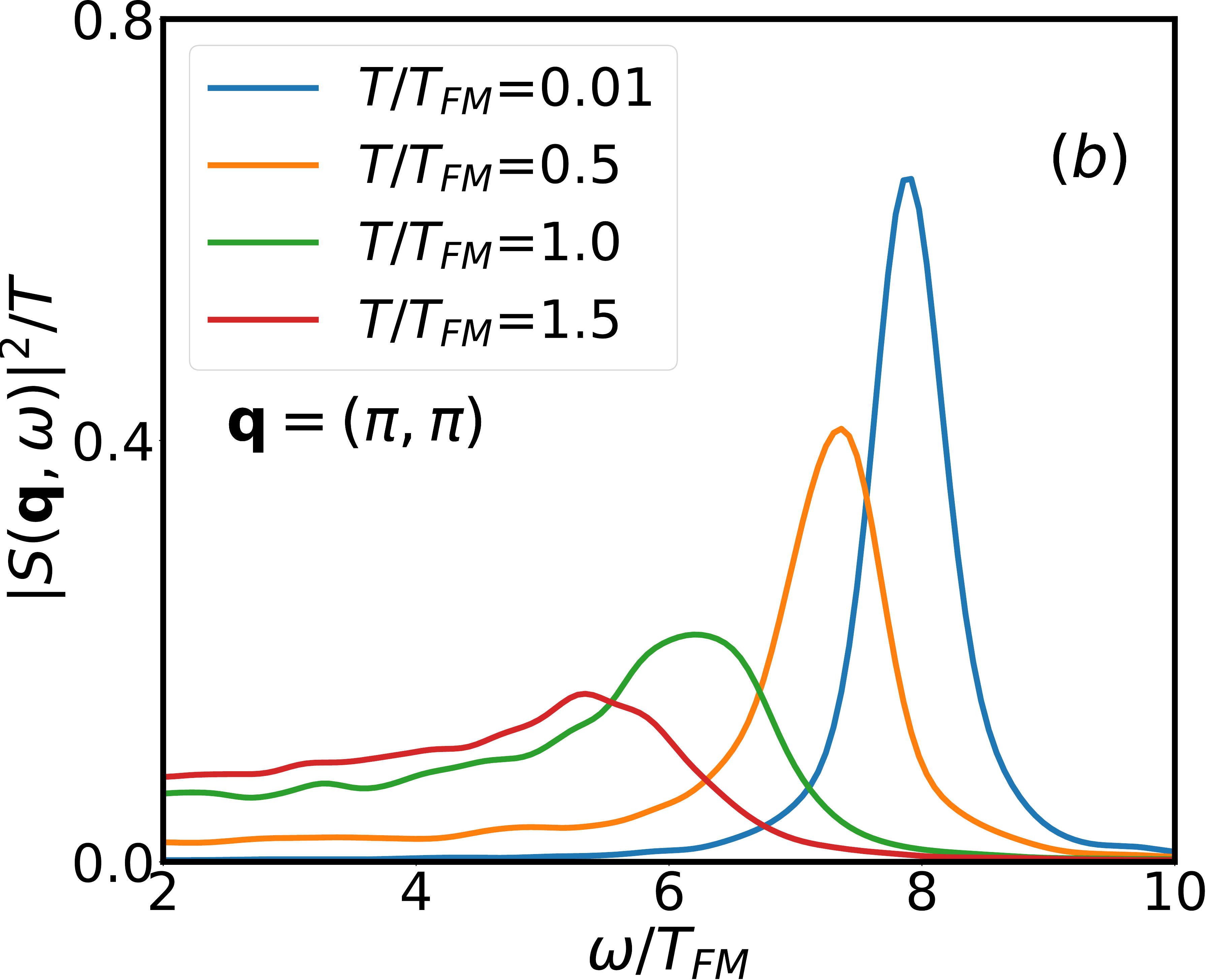}
}
\centerline{
~~~~
\includegraphics[width=4.0cm,height=3.3cm]{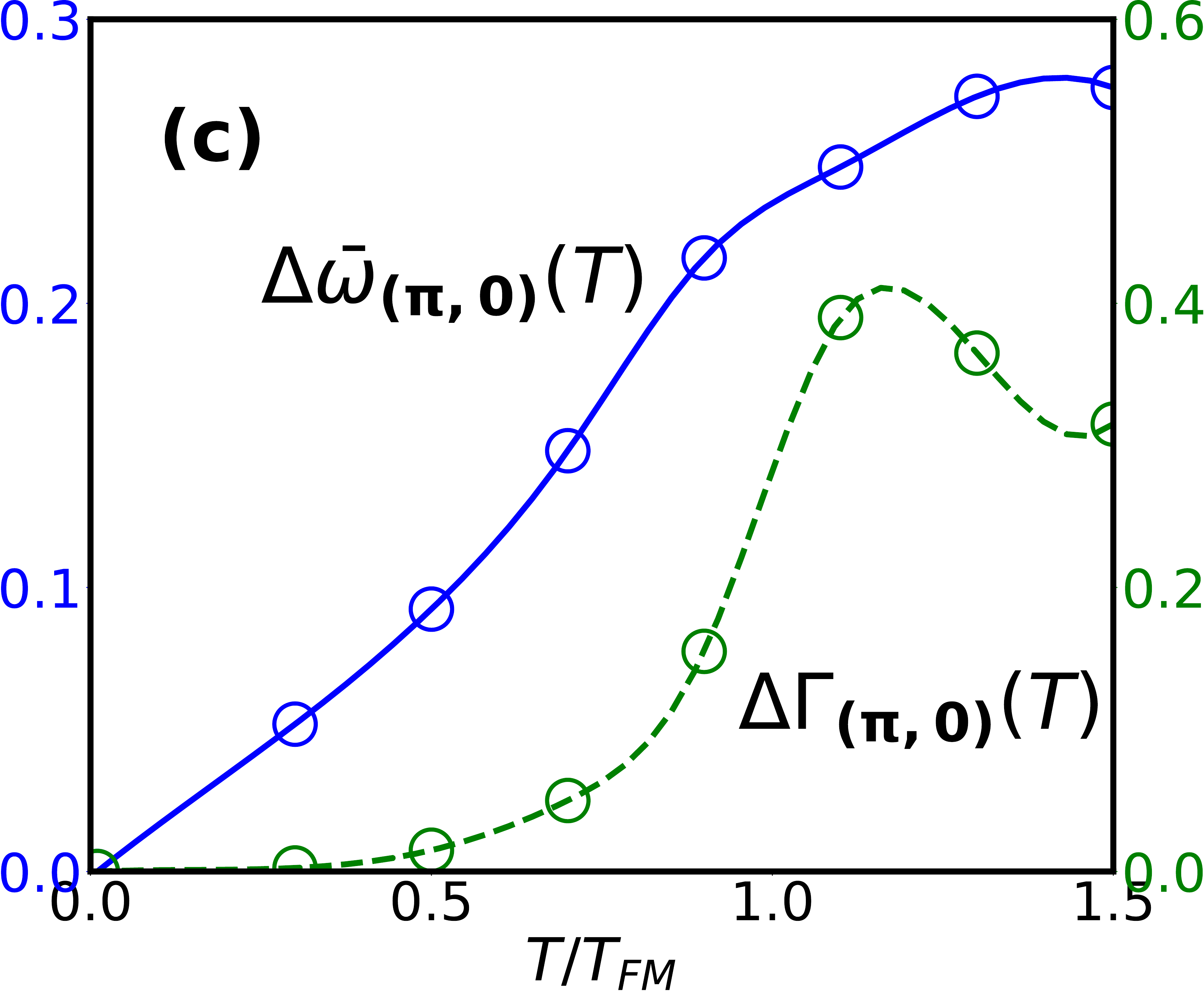}
\includegraphics[width=4.0cm,height=3.3cm]{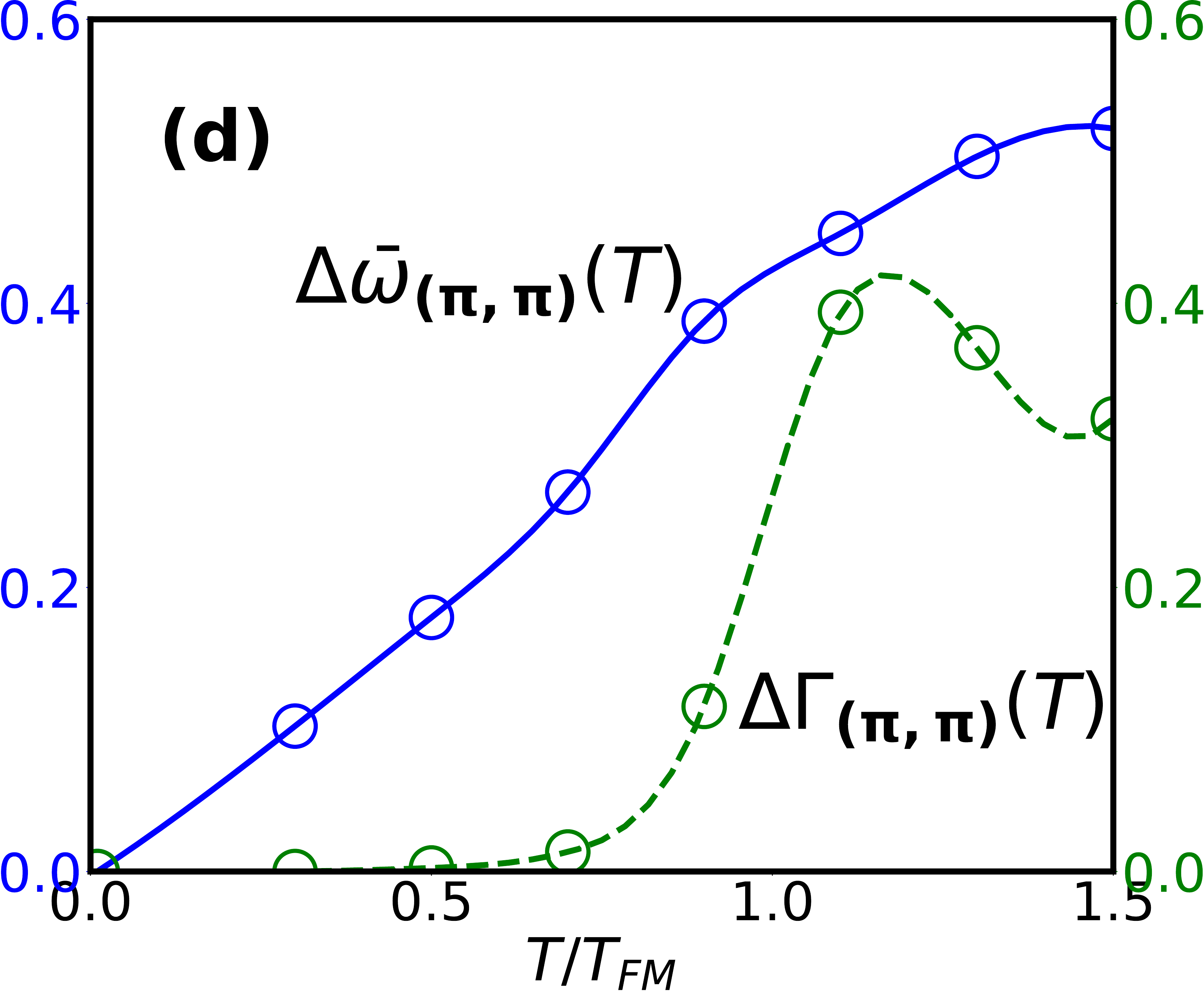}
}
\caption{Top panel: Magnon lineshapes for ${\bf q}=(\pi,0)$ (a)
and ${\bf q}=(\pi,\pi)$ (b) at the same temperatures. The low $T$
spectrum is concordant with a nearest neighbour Heisenberg model for
$J_{eff}\sim0.1t$. Near $T_{FM}$, asymmetric and broad lineshapes are seen.
Bottom: Theoretical estimates of `mode softening' and broadening for
magnons (similar to phonons) for $(\pi,0)$ (c) and $(\pi,\pi)$
(d). We see enhanced softening in the latter.
}
\end{figure}

Our parameter choice was meant to mimic the physics in 
La$_{1-x}$Sr$_x$MnO$_3$
and La$_{1-x}$Ca$_x$MnO$_3$ for $x \sim 0.2-0.3$.
Unlike the real material, the model we use 
is two dimensional, involves Holstein rather than cooperative
JT phonons, and does not include AF couplings. 
As Figs.2(e)-(f) demonstrate the phonon softening at the short
range ordering wavevector follows similar trends in theory
and experiment roughly upto
$T_{FM}$, beyond which they deviate. The fractional softening
near $T_{FM}$ however differs by more than a factor of two. 
Similarly, the thermal
component of phonon broadening, $\Gamma$, has very similar
$T$ dependence in Figs.2(e) and 2(f), but the theory value
is now smaller by about a factor of 2.
What has not been probed experimentally is the signature of
low energy weight at ${\bf q} \sim {\bf q}_{CE}$ at 
$T \lesssim T_c$ - in manganites which show
a thermally induced metal-insulator transition. 
This weight at $\sim 10\%$ of the bare phonon scale is
the key dynamical signature of  short range correlated
large amplitude distortions. Unless ionic 
disorder pins polarons, this 
low energy feature should be visible.

{\bf Conclusions:}
We have presented the first results on the coupled anharmonic
dynamics of phonons and spins that emerges with increasing 
temperature in the Holstein-double exchange model.
Past the low temperature harmonic window, we observe the
expected nonlinearities attributable to `phonon-phonon'
and `magnon-magnon' interactions. Beyond this, however,
we see a striking `two peak' structure in the momentum
resolved phonon spectrum, involving: (i)~low energy
weight at $\omega \ll \Omega$, for ${\bf q} \sim (\pi, \pi)$, 
from slow tunneling of thermally generated spatially
correlated polarons, and (ii)~enhanced damping of  the
high energy, $\omega \sim \Omega$, feature due to 
scattering from magnetic fluctuations. The
magnetic dynamics itself remains mostly insensitive
to the phonon effects and can be described by a Heisenberg
model. Our `high energy' phonon trends compare well with
inelastic neutron scattering in the manganites, although
numerical values differ, and the
low energy features should be visible in the low
disorder samples at $T \gtrsim 0.5T_{FM}$.

\bibliographystyle{unsrt}

\begin{thebibliography}{99}
\bibitem{crossover} There is no genuine ferro to para transition in
an $O(3)$ model in 2D. However, the correlation length grows
exponentially as $T$ is reduced below the effective exchange scale,
$J_{eff}$, and in any finite size calculation it is difficult to
distinguish a merely long range `correlated' state from a genuinely
ordered one.
\bibitem{expt-theory}
The quantitative discrepancy between our results and 
experiments may arise partly from the use of acoustic phonons
in the experiments, although these phonons had the same
$\Sigma_{3}$ symmetry as JT displacements.
\end{thebibliography}

\end{document}


\author{Sauri Bhattacharyya$^{1}$, Sankha Subhra Bakshi$^{1}$, 
	Saurabh Pradhan$^{2}$ and Pinaki Majumdar$^{1}$}

\affiliation{
$^1$~Harish-Chandra Research Institute, HBNI,
Chhatnag Road, Jhunsi, Allahabad 211 019, India\\
$^2$~Department of Physics and Astronomy, Uppsala University,
751 05 Uppsala, Sweden }

\title{Supplemental Material for
``Strongly anharmonic collective modes in a coupled 
electron-phonon-spin problem''}

\date{\today}

\maketitle

We first discuss the derivation of the respective 
Langevin equations for phonons and spins. Next, the
consistency between the results obtained using
the present scheme and exact diagonalization based
Monte Carlo is shown. Then, to hint at the various 
thermal regimes in phonons, the distribution of 
displacement and density fields are featured.
A comparison of spectra in the present model
with that obtained from a pure Holstein and a
pure double exchange model are depicted afterwards.
This elucidates the role of magnon-phonon coupling.
Finally, real-time trajectories of phonon and
spin fields are exhibited to create an insight
into the detailed results.

\section{Deriving the Langevin equation}

\noindent
We reproduce the evolution equation below for reference
\begin{eqnarray}
M{ {d^2x_i} \over dt} & = &
-D_{ph} { {dx_i} \over dt}   - K x_i -
{ {\partial {\langle H_{el} \rangle }}   \over {\partial x_i }}
+ \xi_i(t) \cr \cr
{{d {\vec S}_i} \over {dt} }~~ & =&
-\vec{S}_{i}\times({{\partial
{\langle H_{el} \rangle }}\over {\partial \vec{S}_i }}
+ \vec{h}_i) + ~D_{s}\vec{S}_{i}\times (\vec{S}_{i}
\times{{\partial {\langle H_{el} \rangle }}\over
{\partial \vec{S}_{i} }} ) \cr
\cr
H_{el}~ & = & \sum_{ij}(t_{ij} - \mu \delta_{ij})
\gamma^{\dagger}_{i}\gamma_{j} - g\sum_in_ix_i
\cr
t_{ij}/t ~&= & \sqrt{(1+\vec{S}_i.\vec{S}_j)/2}
\nonumber
\end{eqnarray}
In what follows we comment on (i)~the origin of the
phonon and spin equations, (ii)~the `noise', $\xi_i(t)$
and ${\vec h}_i(t)$,  (iii)~the choice of the damping 
parameters $D_{ph}$ and $D_{s}$, and (iv)~the robustness of the 
`equal time' properties (at equilibrium) to variation in the 
damping parameters.

\subsection{Phonon equation}

The derivation of the phonon equation of motion has been outlined
before \cite{sauri} in case of the pure Holstein model, 
which is in turn motivated by earlier work \cite{martin}. 
To quickly recapitulate, 
(i)~A real time path integral is set up in the Keldysh formalism in 
terms of coherent state fields corresponding to $x_{i}$ and $c_{i}$ 
operators. (ii)~Integrating out the quadratic electrons we obtain
an effective action for the phonon fields on the Keldysh contour. 
(iii)~We transform to `classical' and `quantum' variables, the average
and difference of phonon fields on the forward and backward branches
of the Keldysh contour, and 
expand the action perturbatively in the `quantum' fields, assuming 
that the phonon timescale, ${\Omega}^{-1}$, is much larger than 
the electronic timescale, $t^{-1}$ ($t$ is the hopping).
(iv)~Integrating out the `quantum' fields at Gaussian level 
leads to a stochastic equation of motion for the `classical' fields.

The main difference here 
with respect to the pure Holstein model is the
(time dependent) modulation of the hopping amplitude, $t_{ij}$,
due to fluctuation in the orientation of the spins. 
The implicit dependence of the `force' term  
${ {\partial {\langle H_{el} \rangle }}   \over {\partial x_i }}
= gn_i(t)$ on the spin configuration is relatively weak.

\subsection{Spin equation}

The spin equation is written based on arguments presented in 
Ref.4. The significance of the different terms is as follows:
(i)~The torque term describes semiclassical 
dynamics of the fixed magnitude moments in an effective magnetic field. 
In our case, this
field is computed as a gradient of the instantaneous  electronic energy
with respect to the local spin ${\vec S}_i$. 
(ii)~The dissipation is proportional to the angular 
momentum of the moments, in analogy with the motion of ions.
When the
the dynamics is restricted to the sufrace of a unit sphere, 
our LLGB equation emerges. 

There are alternate derivations starting from a model of 
local moments coupled to phonons \cite{rebei1} and/or conduction 
electrons \cite{rebei2,mera} and then employing the Keldysh formalism. 
One makes similar assumptions there as in the electron-phonon case, 
namely that the moments are `slow' compared to other microscopic 
degrees of freedom and that
the electronic density of states is gapless and 
conforms to the Ohmic dissipation model.

The spin equation differs from that for the double exchange
model only in the implicit dependence of bond kinetic energy
average $\langle\gamma^{\dagger}_i \gamma_j\rangle$ on the phonon
field. It differs from the Heisenberg model, additionally,
via the factor $1/\sqrt{(1 + {\vec S}_i.{\vec S}_j)/2}$,
which modifies the local field acting on the site $i$.

\subsection{Specifying the noise}

The `noise' which drives the phonon field has a correlator
$$\langle \xi_i(t) \xi_j(t') \rangle = g^{2}\Pi_{ij}(t,t^{\prime})
$$
where $\Pi$ is the Keldysh component of
the electronic polarizability, in the background $x_i(t)$.
On the assumption that we are at `high
temperature', $\Omega \ll k_BT$, and that
the electronic density of states is gapless,
we obtain the Langevin form used in the paper.
Although we use the high temperature
approximation throughout to simplify the noise correlator,
the essential dynamical features of the phonons
are retained even for
$k_{B}T \lesssim \Omega$ through the
`electronic force' term.

The spin noise ${\vec h}_i$ 
is chosen to be Markovian, obeying the 
fluctuation-dissipation theorem, for simplicity.
In an actual derivation from a Keldysh action, 
the noise correlator is proportional to the Keldysh spin
polarizability $[\Pi_{ij}^{S}(t,t^\prime)]^{\mu\nu}$, 
that has a natural frequency scale $\sim t$. For low
frequencies ($\omega \ll t$) and high temperatures 
($\omega \ll k_{B}T$), the present Markovian form may
be used. Once again, we apply this approximation at all
temperatures, assuming that the key dynamical features
are captured through the `electronic torque' term.

\begin{figure}[t]
\centerline{
\includegraphics[width=4.1cm,height=3.7cm]{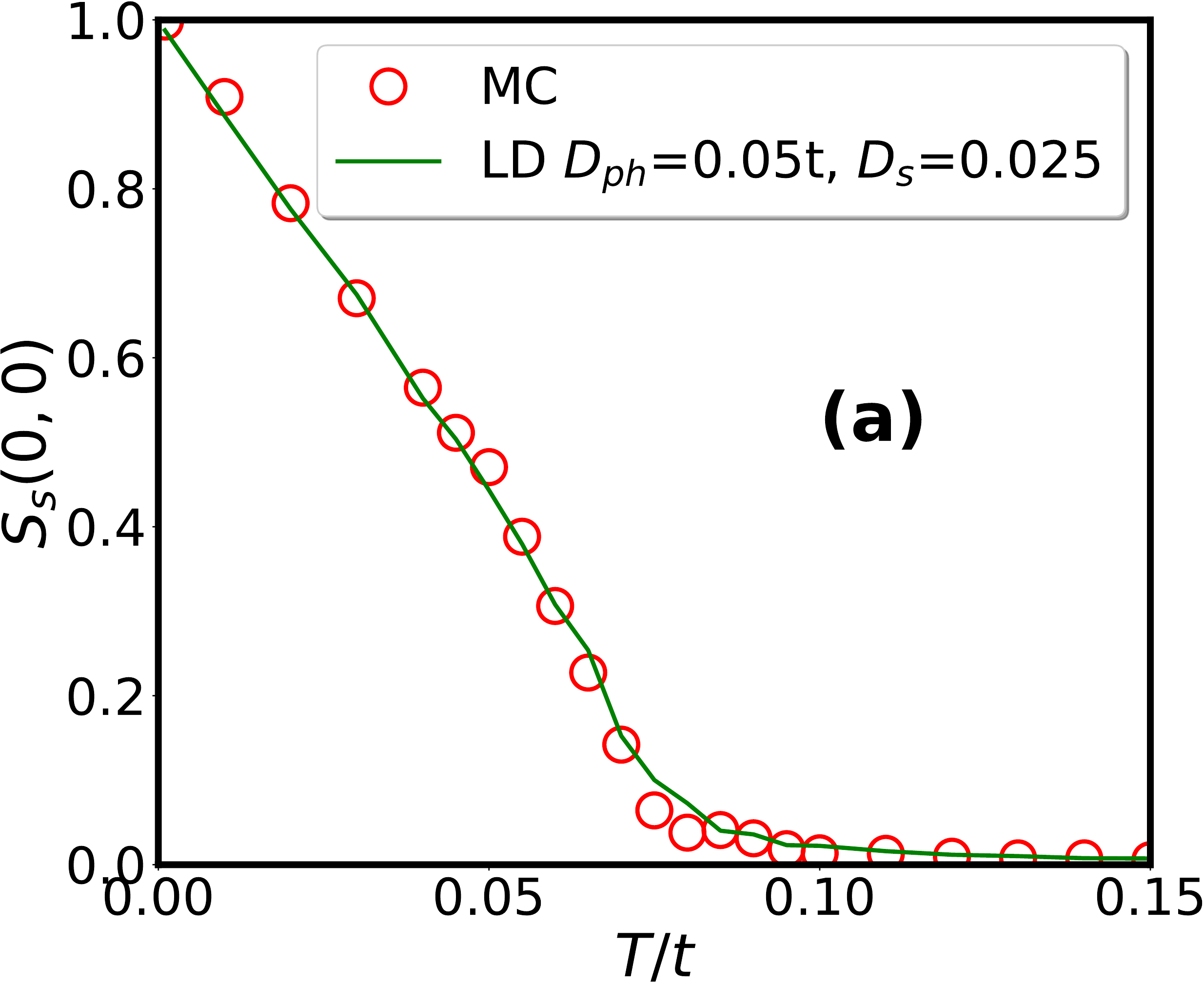}
\includegraphics[width=4.1cm,height=3.7cm]{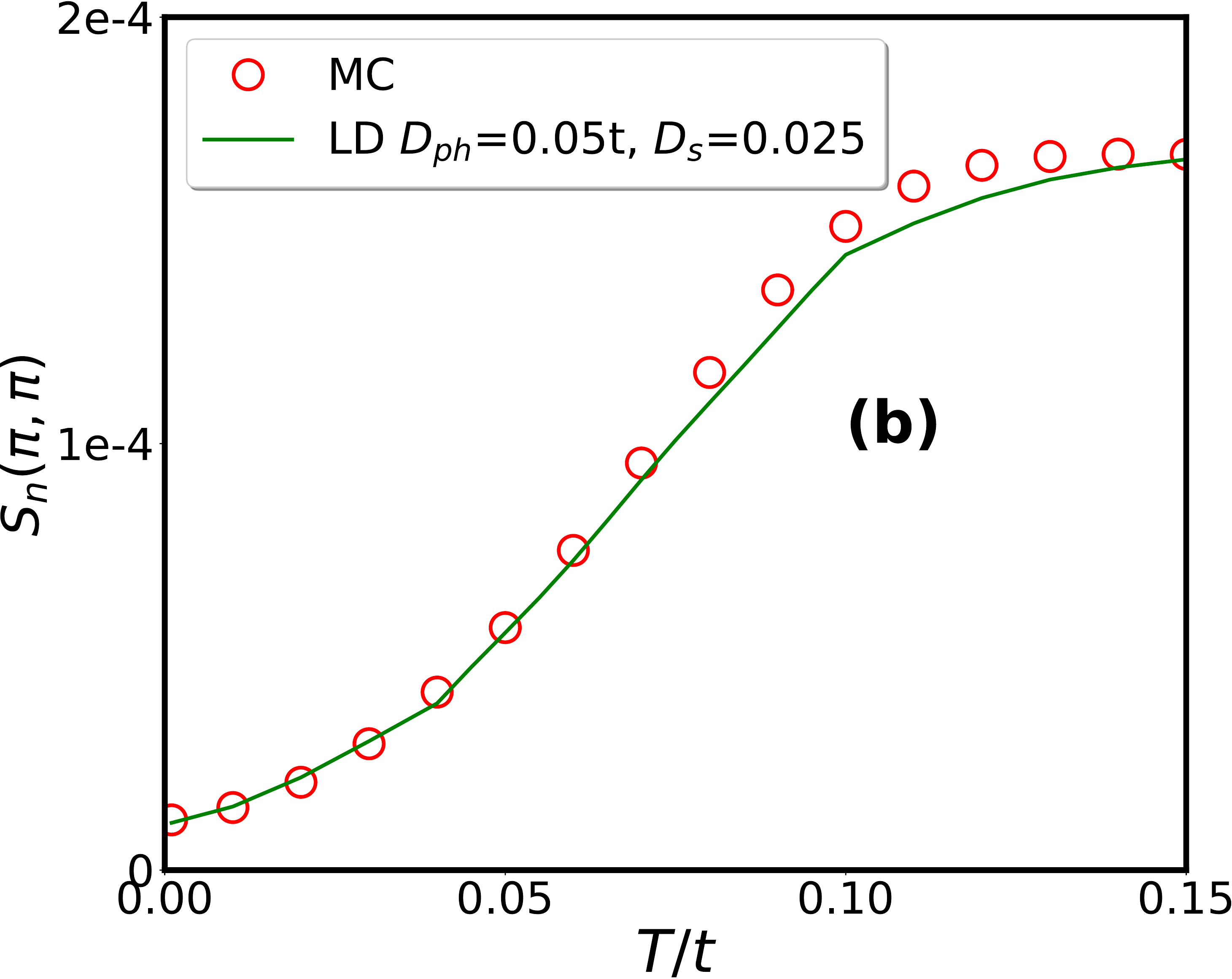}
}
\centerline{
\includegraphics[width=4.1cm,height=3.7cm]{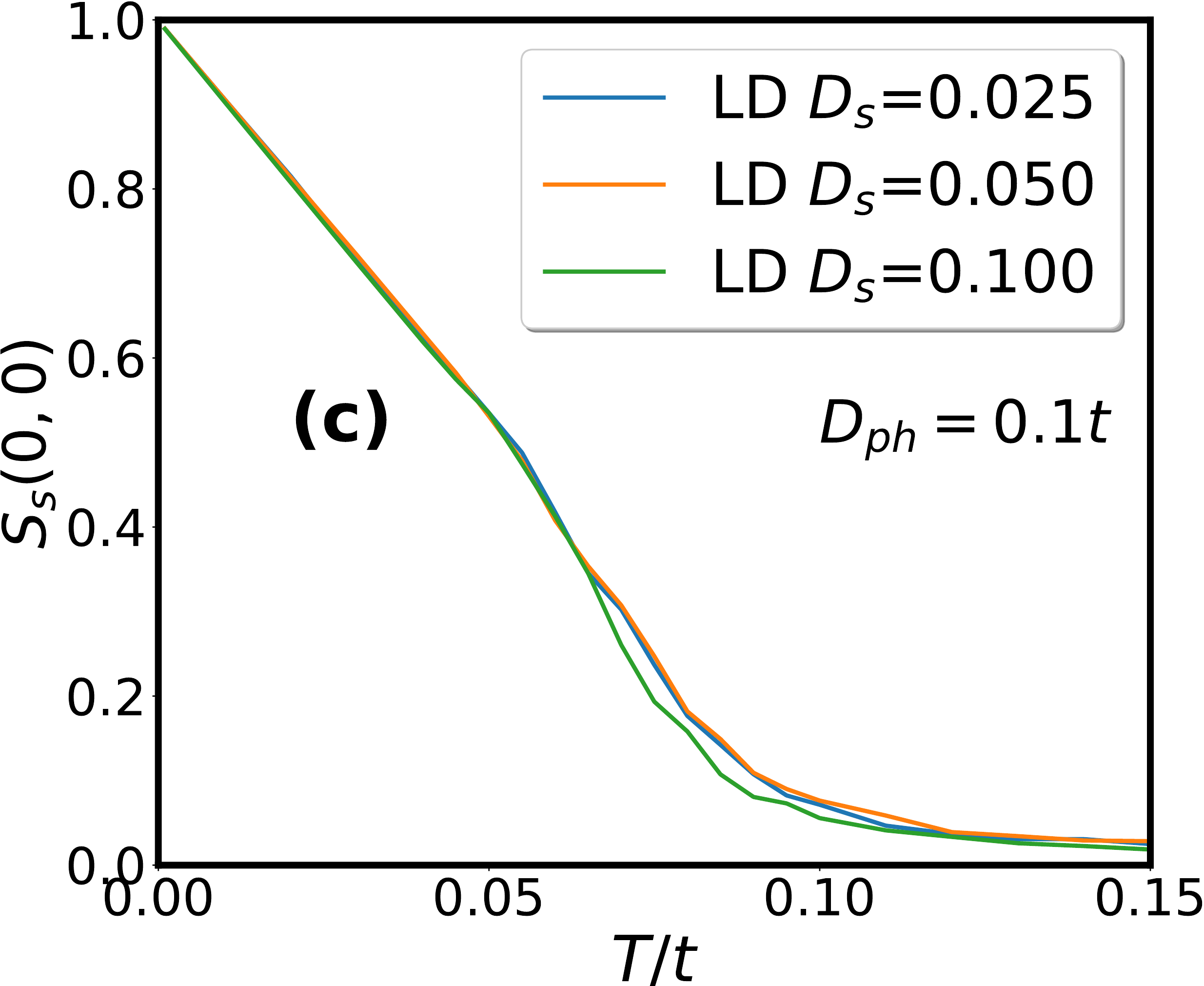}
\includegraphics[width=4.1cm,height=3.7cm]{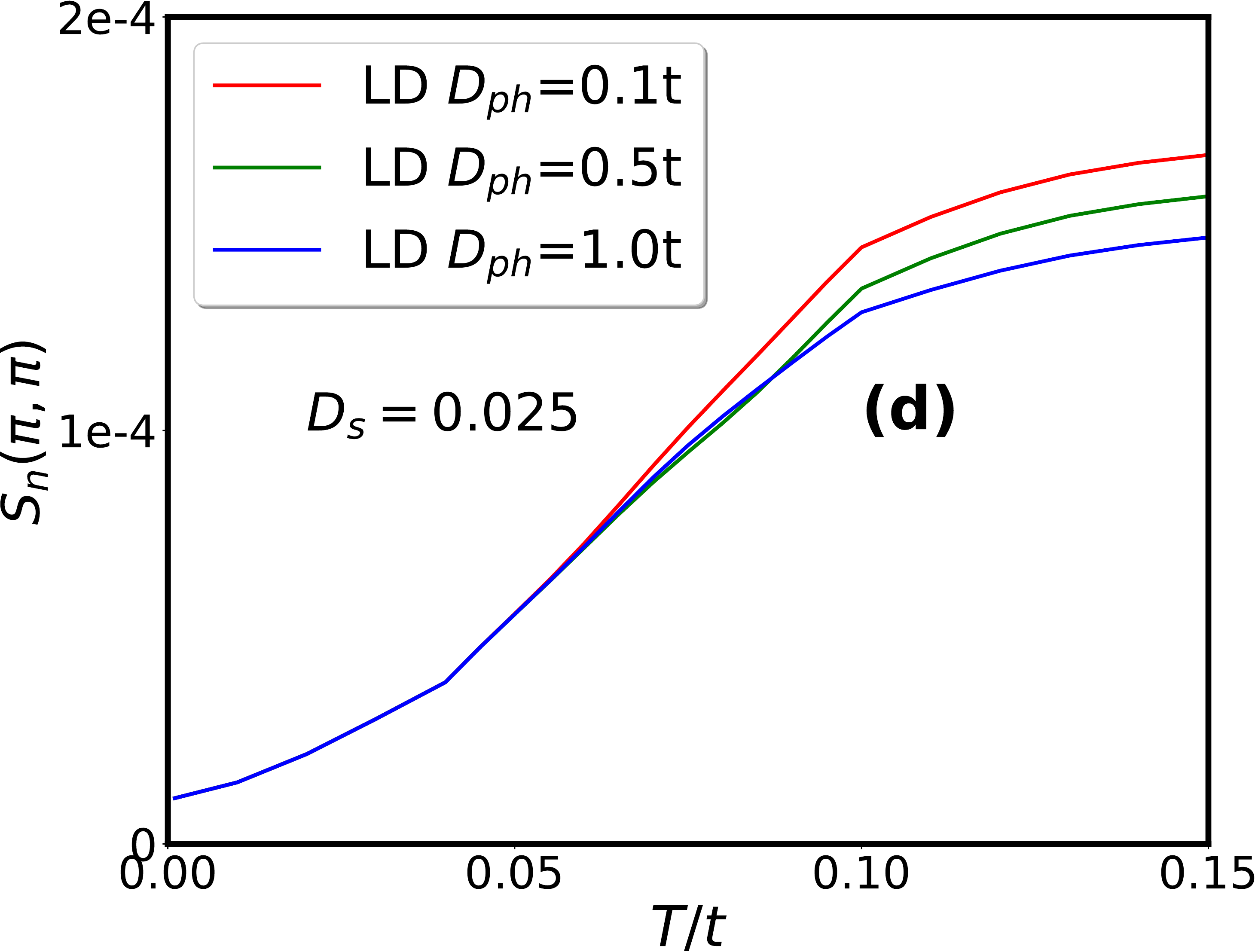} }
\caption{(a)-(b): Comparison of Langevin results with Monte Carlo.
(a)~The ferromagnetic feature $S_{s}(0,0)$ in
the magnetic structure factor and (b)~the Fourier transform
of the density-density correlation $S_{n}$ at ${\bf q} = (\pi,\pi)$.
There is a very good agreement for the chosen $D_s$ and $D_{ph}$ values.
(c)~The dependence of $S_{s}(0,0)$
on $D_s$ for three values $D_{s}={0.025,0.05,0.1}$, and
(d)~of $S_{n}(\pi,\pi)$ on $D_{ph}$ for three values $D_{ph}={0.1t,0.5t,1.0t}$.
In both cases, we see an insensitivity to dissipation rates.
} 
\end{figure}

\subsection{The damping parameters}

Microscopically, in the Holstein double exchange model, the phonons
couple to the local density and the moments to the local conduction
electron spin. Hence, the dissipation channels arise from 
particle-hole excitations in the charge and spin sectors.
These involve the charge
and spin susceptibilities of the parent electronic system. 
At low temperature
one may approximate these by their `bare' versions. 
The dissipation rates $D_{s}$ and $D_{ph}$ arise from the imaginary parts 
of the `linear', low-frequency part of these functions. 
$D_{ph}$ carries a prefactor of $g^{2}$, while $D_s$ is proportional
to $t^{2}$. An actual calculation reveals that $D_{s}\sim0.1t$ 
and $D_{s}\sim0.1$ at low temperatures.

\section{Consistency between Langevin dynamics and Monte Carlo}

The `static' physics at equilibrium for the Holstein-double exchange
model can be extracted from an exact diagonalisation based Monte Carlo
(ED-MC), when we assume the phonons and spins to be classical. That
assumes $M \rightarrow \infty$, {\it i.e}, $\Omega \rightarrow 0$,
and the `size' of the spin $S \rightarrow \infty$ (to be absorbed 
in the Hund's coupling). In that case
the probability for a phonon-spin configuration is:
\begin{eqnarray}
P\{x, S\} &\propto&  Tre^{- \beta (H_{el} + H_{stiff})} \cr
\cr
H_{el}~~ & = & \sum_{ij}(t_{ij} - \mu \delta_{ij})
\gamma^{\dagger}_{i}\gamma_{j} - g\sum_in_ix_i
\cr
t_{ij}/t ~&= & \sqrt{(1+\vec{S}_i.\vec{S}_j)/2}
\cr
\cr
H_{stiff} &=& {1 \over 2} K\sum_i x_i^2
\nonumber
\end{eqnarray}

\begin{figure}[b]
\centerline{	
\includegraphics[width=3.7cm,height=3.7cm]{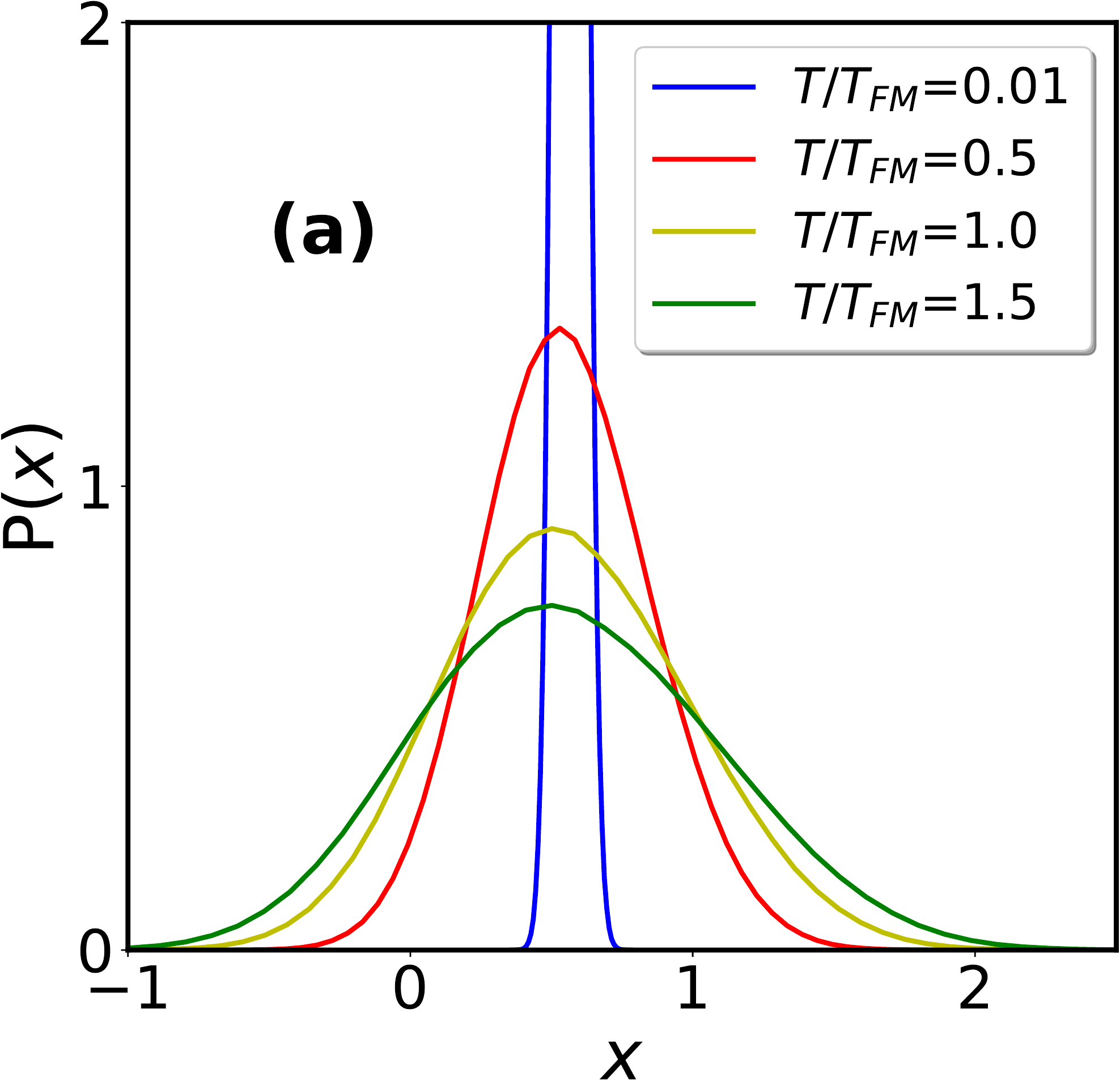}
\includegraphics[width=3.7cm,height=3.7cm]{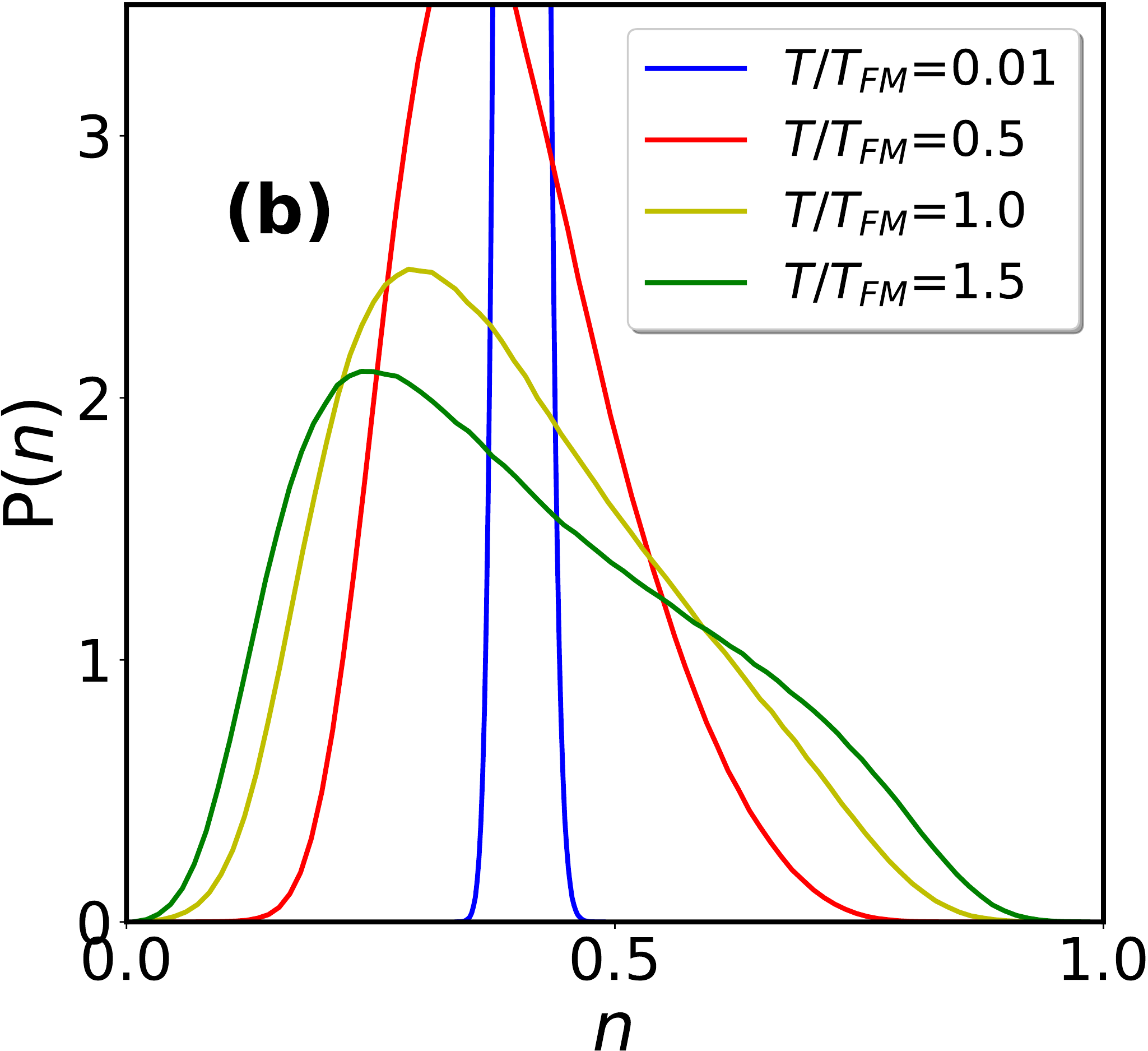}
}
\caption{(a)-(b): The distribution of distortions ($P(x)$) 
and density ($P(n)$) for four temperatures- 
$T/T_{FM}={0.01,0.5,1.0,1.5}$. The former 
broadens gradually on heating up.
The latter shows a more interesting evolution from 
a sharp unimodal profile ($T<<T_{FM}$) through a positively 
skewed, long tailed distribution (near $T_{FM}$) to finally
a plateau-like structure beyond $T_{FM}$. The quick onset of 
anharmonicity, presence of large distortions near $T_{FM}$, 
and the merging of large oscillations with polarons at
high $T$, are all hinted at.
}
\end{figure}

\begin{figure}[b]
\centerline{
\includegraphics[width=4.1cm,height=3.7cm]{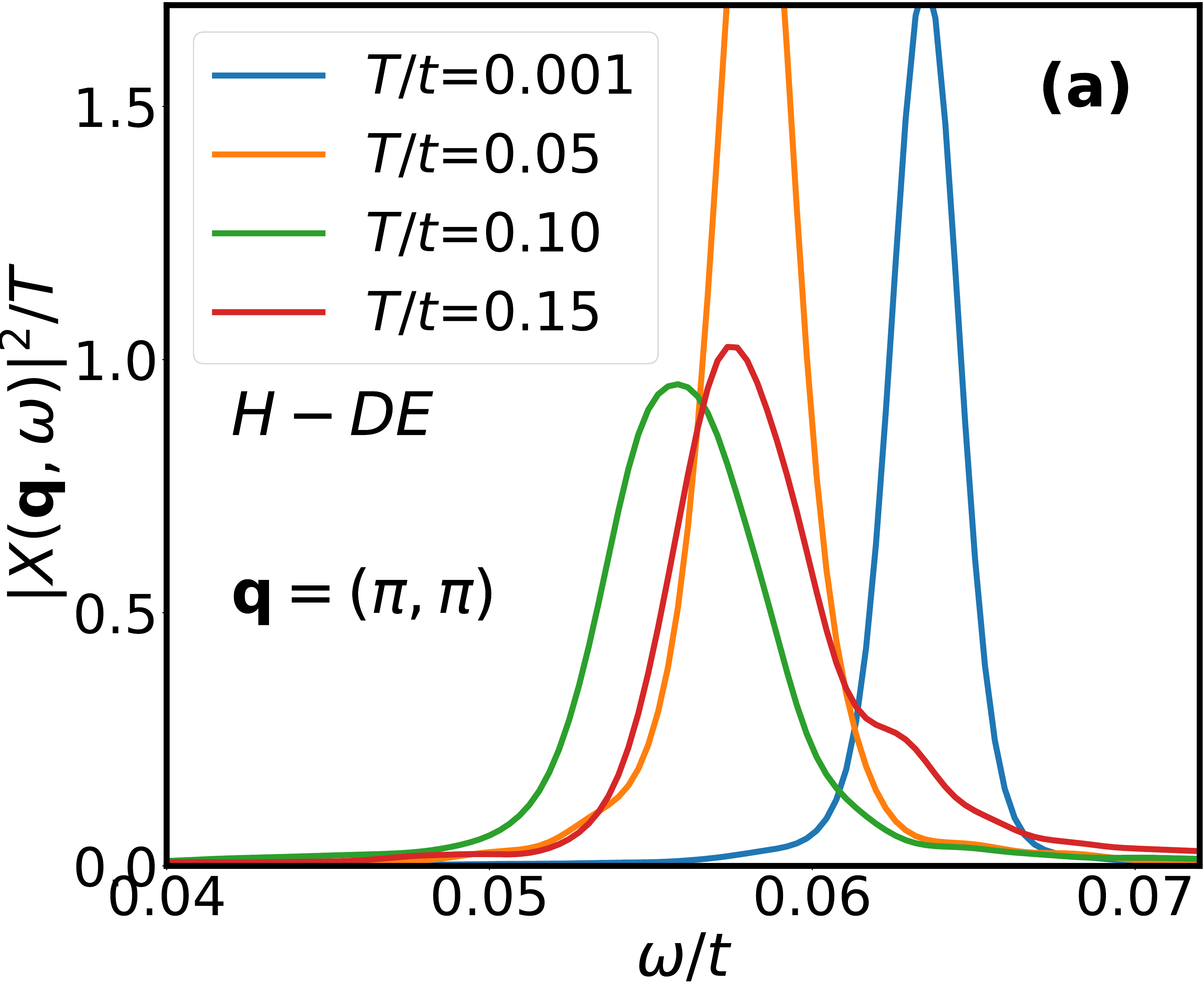}
\includegraphics[width=4.1cm,height=3.7cm]{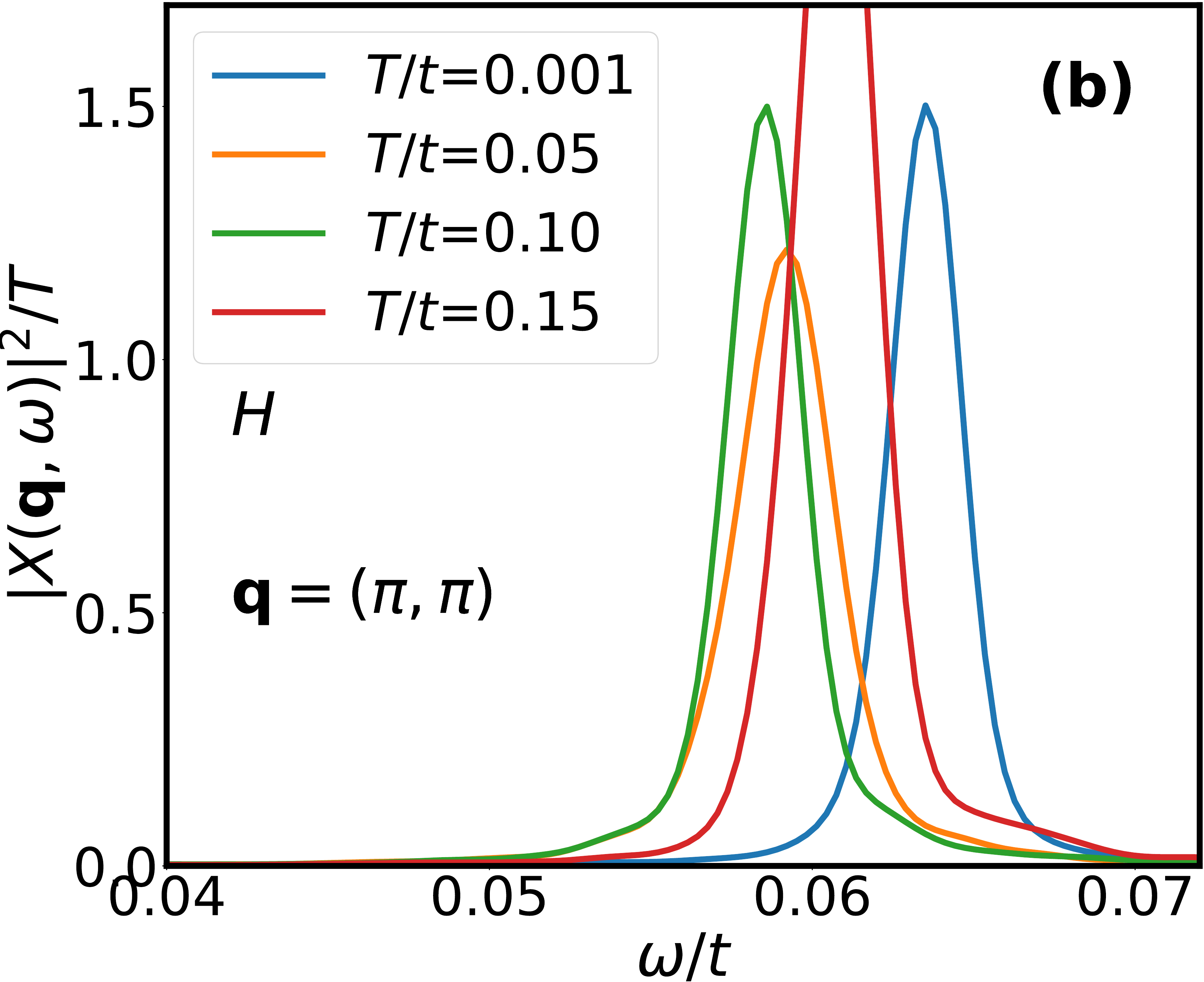}
}
\centerline{
\includegraphics[width=4.1cm,height=3.7cm]{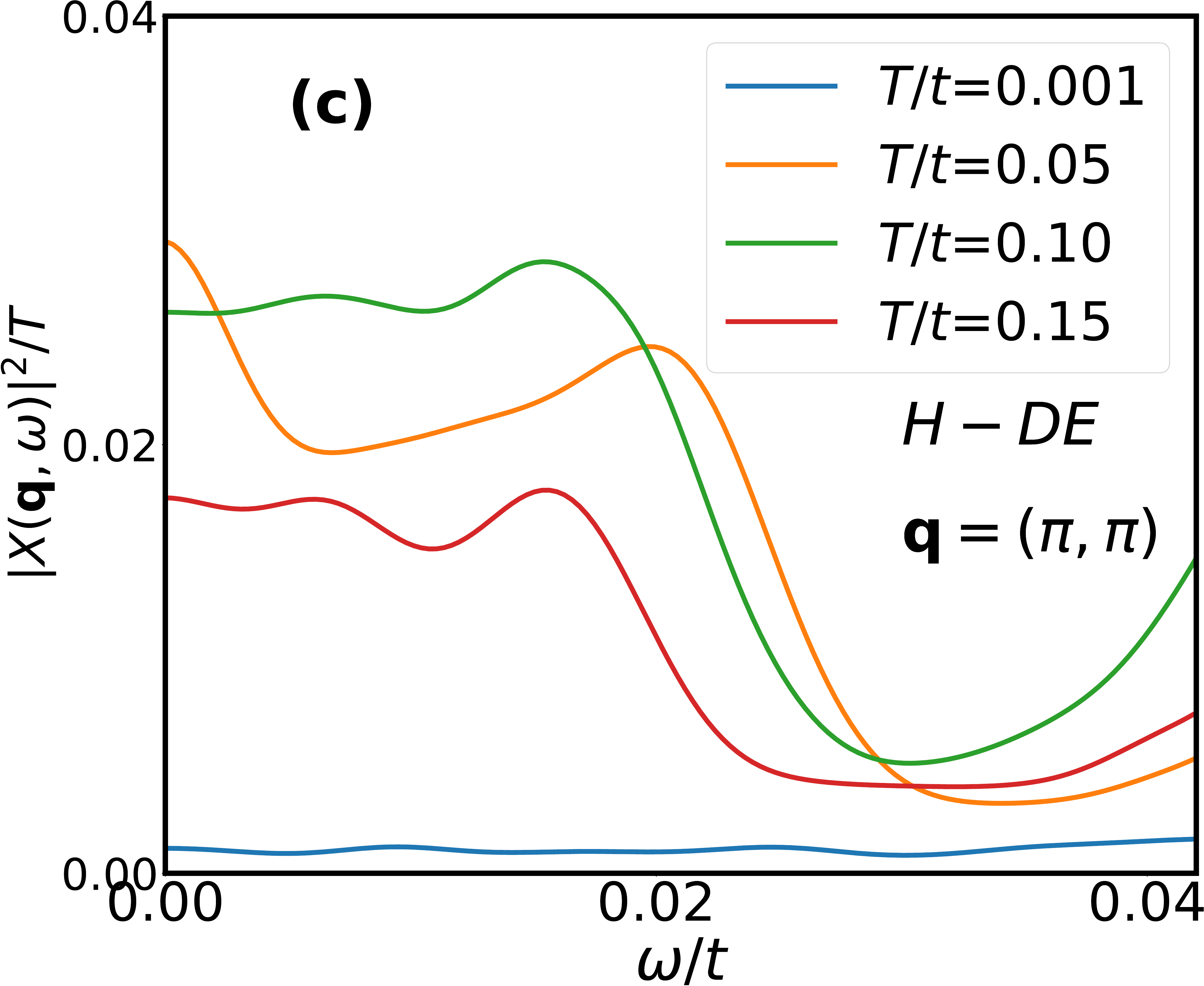}
\includegraphics[width=4.1cm,height=3.7cm]{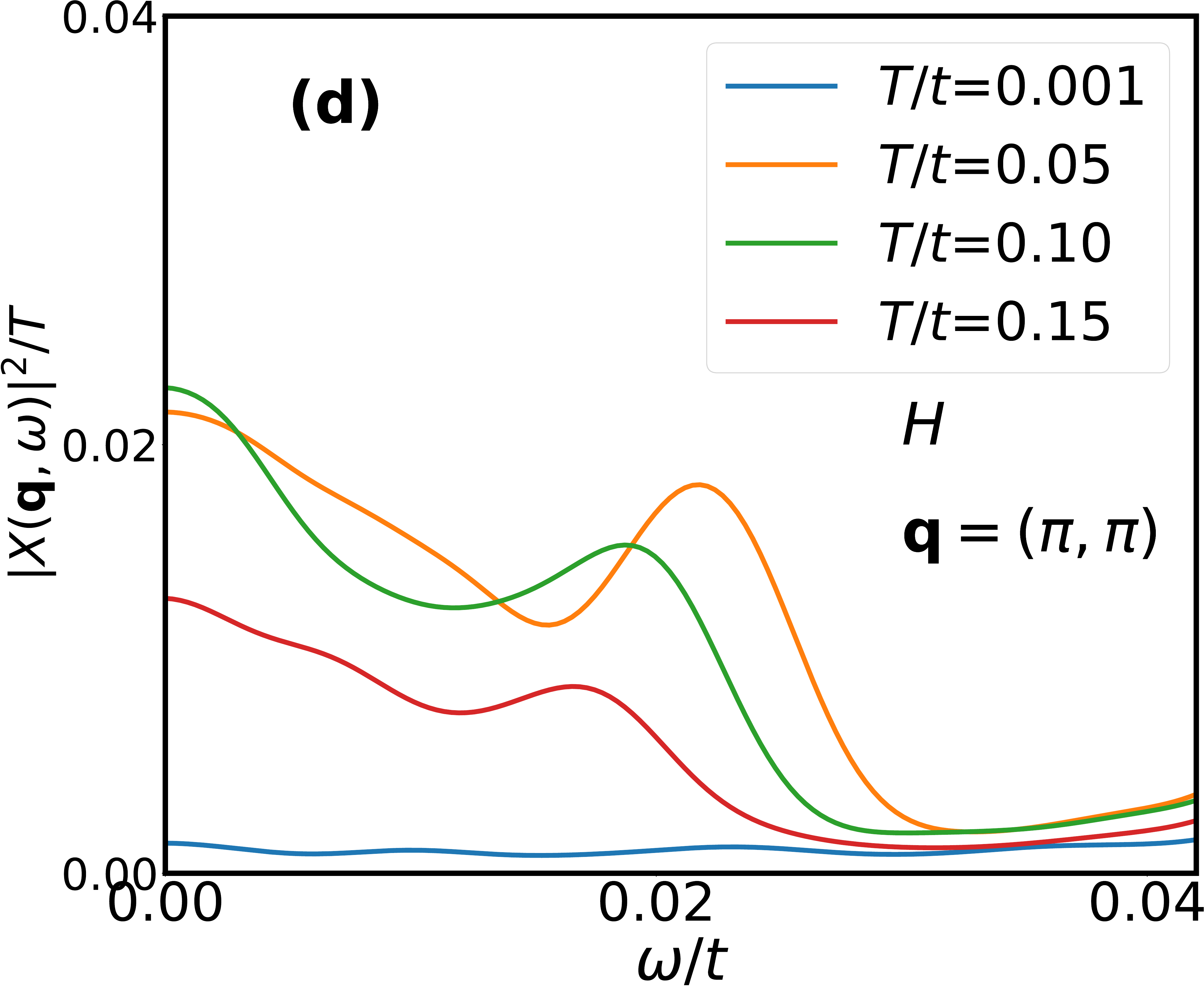}
}~~
\centerline{
\includegraphics[width=4.1cm,height=3.7cm]{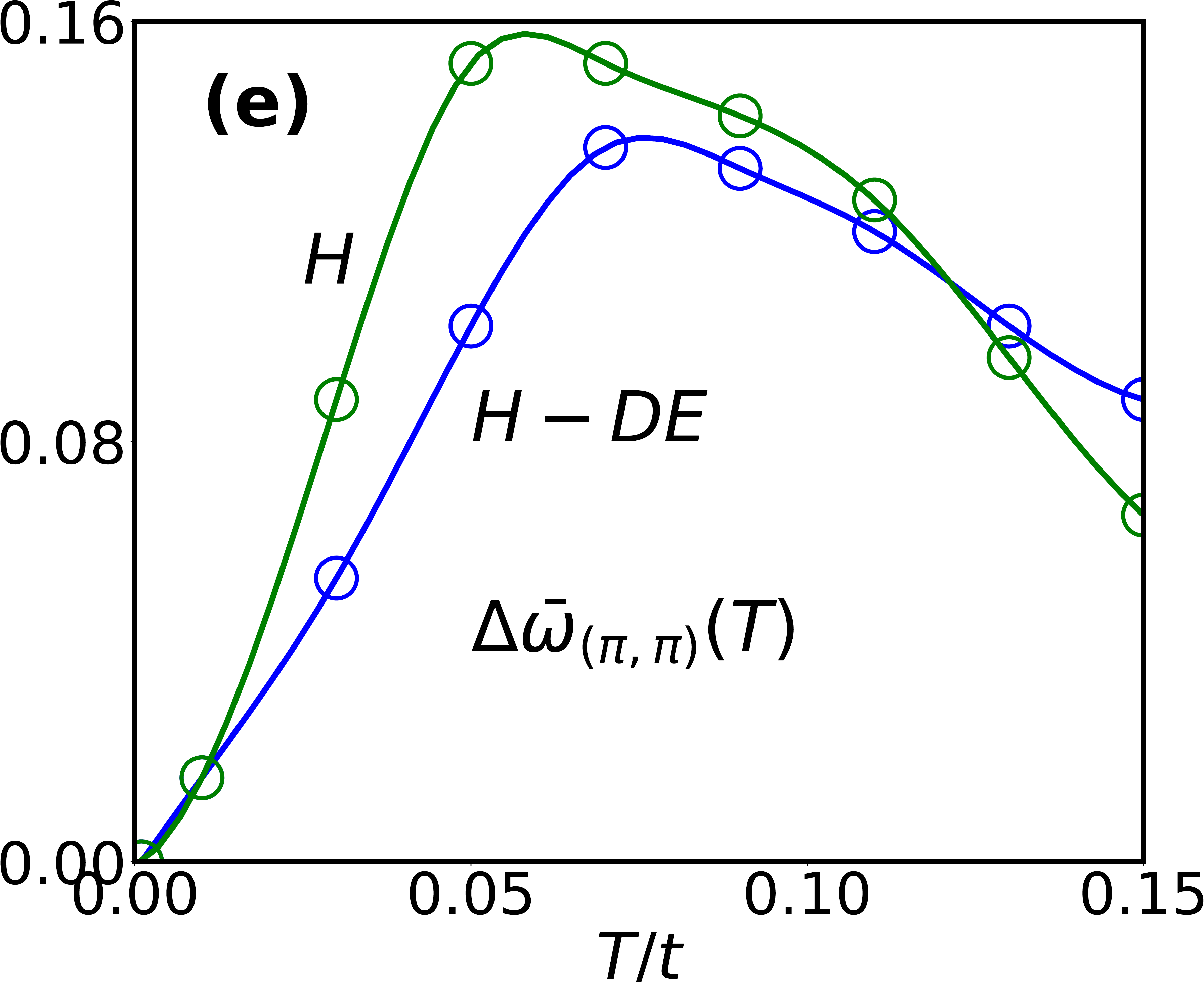}
\includegraphics[width=4.1cm,height=3.7cm]{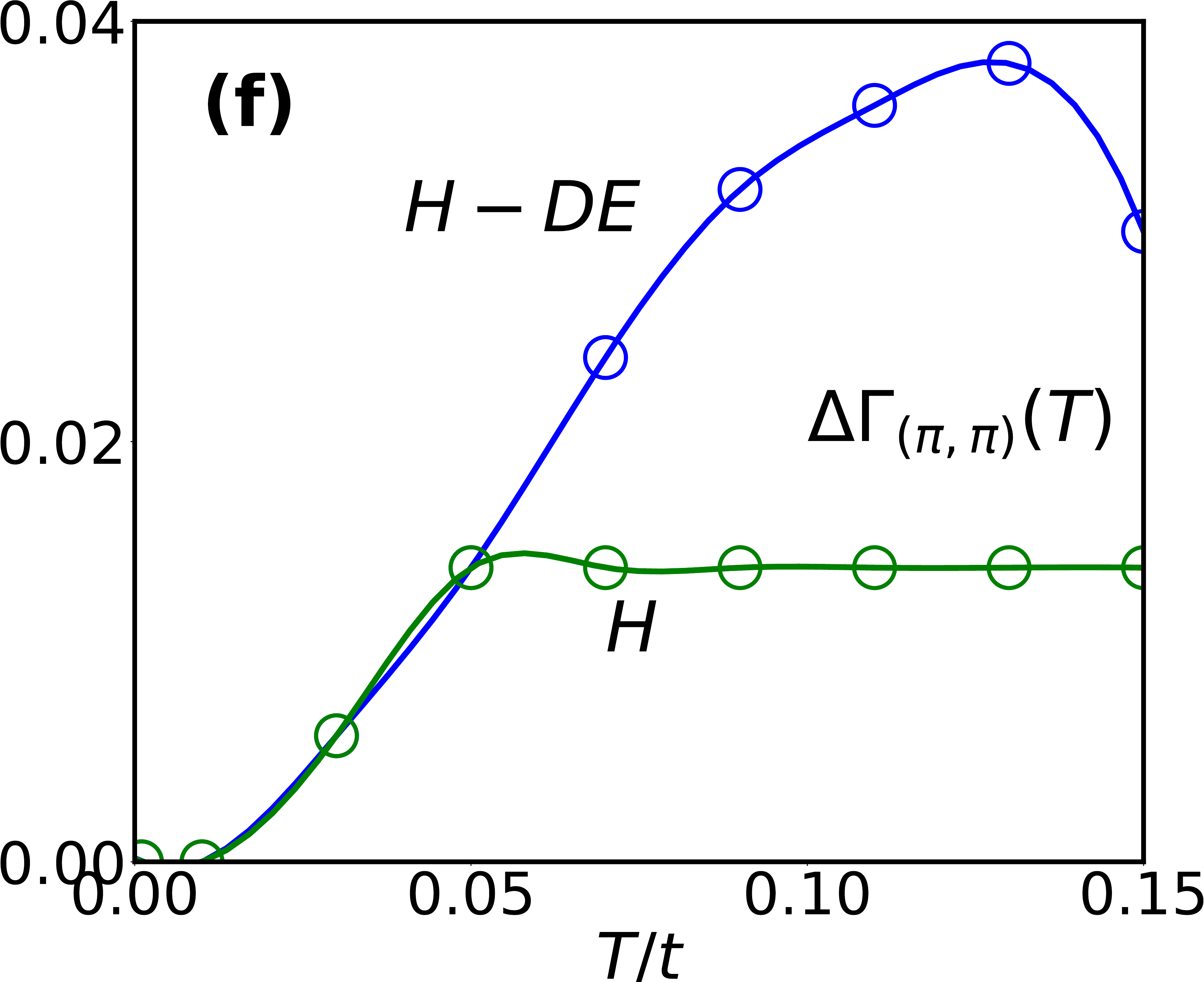}
}
\caption{(a)-(b): Comparison of `high energy' part of 
phonon lineshapes at ${\bf q}=(\pi,\pi)$ for different 
temperature regimes between the present model (H-DE) and a 
pure Holstein model (H) at comparable density. 
(c)-(d): Same for `low energy' part of the spectra.
Similar qualitative features are seen in both cases,
but the effects are stronger in the former.
(e): The extracted softening
$\Delta\bar{\omega}_{(\pi,\pi)}(T)$ compared between 
the H-DE and Holstein models. The results are broadly 
similar.   
(f): The linewidth
$\Delta\Gamma_{(\pi,\pi)}(T)$ 
compared between the H-DE and Holstein models.
In the Holstein model the width saturates at some temperature, 
without the peak that one observes in the H-DE model. The larger 
H-DE damping and the peak feature in $T$ dependence arise due
to the magnetic degrees of freedom.
}
\end{figure}
The trace over the electrons can be performed numerically by diagonalising
$H_{el}$. This forms the heart of ED based MC and yields results that
do not have the effect of the parameters 
$M,~D_{ph},~D_{s}$ that enter the Langevin equation.
It is important to verify that the Langevin approach, with its additional
dissipative parameters, yields equilibrium 
results that are consistent with MC.

In Fig.1, we show the behaviour of the spin structure 
factor $S_s({\bf q})$ and the density structure factor
$S_n({\bf q})$ at ${\bf q} = (0,0)$ and
$(\pi, \pi)$ respectively.
\begin{eqnarray}
S_s({\bf q}) &=& {1 \over N^2} \sum_{ij} {\vec S}_i.{\vec S}_j
e^{i {\bf q}.({\vec r}_i - {\vec r}_j)} \cr
S_n({\bf q}) &=& {1 \over N^2} \sum_{ij} \langle n_i \rangle
\langle n_j \rangle 
e^{i {\bf q}.({\vec r}_i - {\vec r}_j)} 
\nonumber
\end{eqnarray}
We test out  different $D_s$ and $D_{ph}$ values. 
In the top panel, the
results on these obtained using the two methods
(Langevin dynamics and Monte Carlo annealing) are shown. 
The match is remarkable.
For both the spin and density cases, the  bottom panel shows
an insensitivity of the Langevin inferred structure factor
to the damping rates $D_{ph}$ and $D_s$.

\begin{figure}[b]
\centerline{
\includegraphics[width=4.1cm,height=3.7cm]{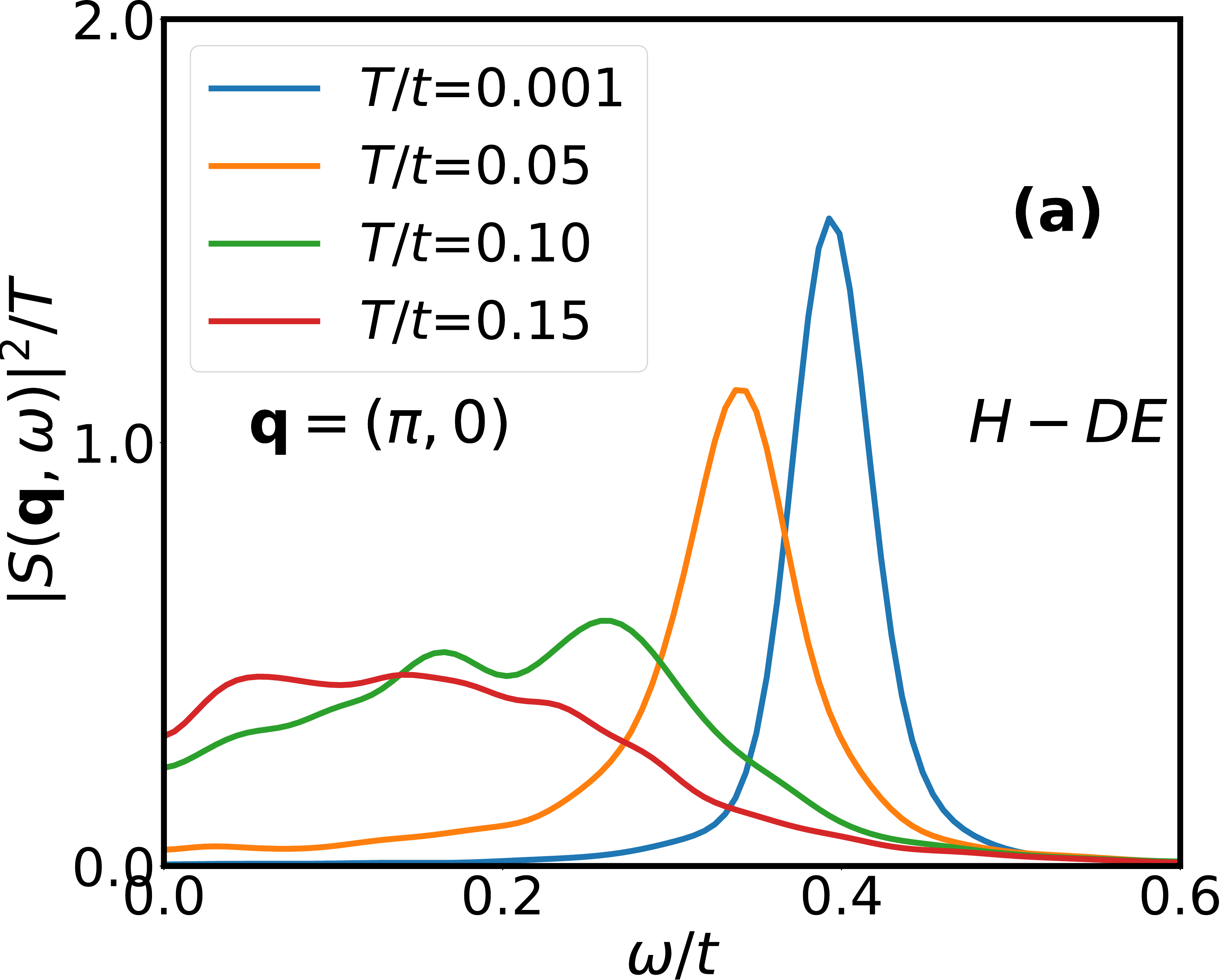}
\includegraphics[width=4.1cm,height=3.7cm]{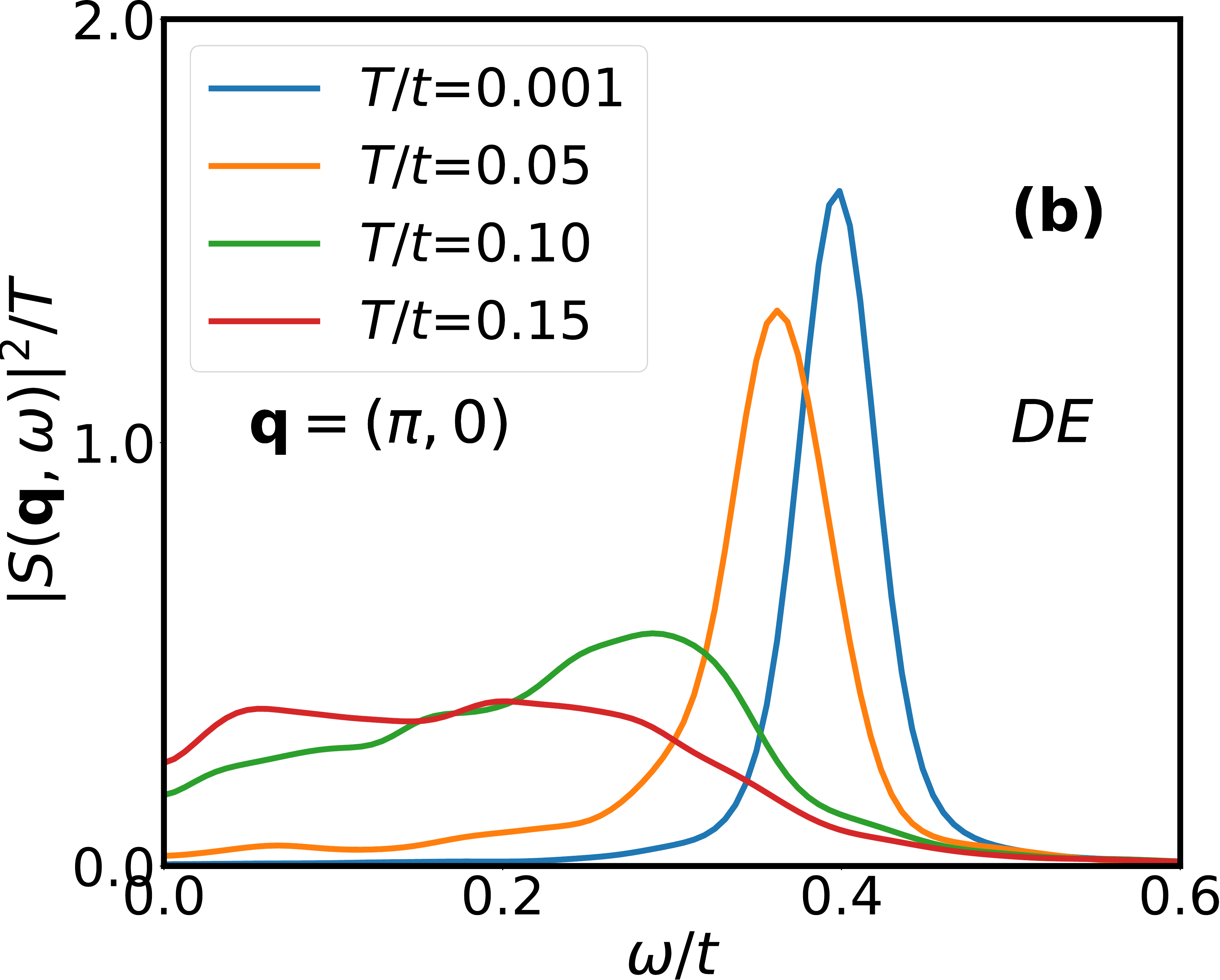}
}
\centerline{
\includegraphics[width=4.1cm,height=3.7cm]{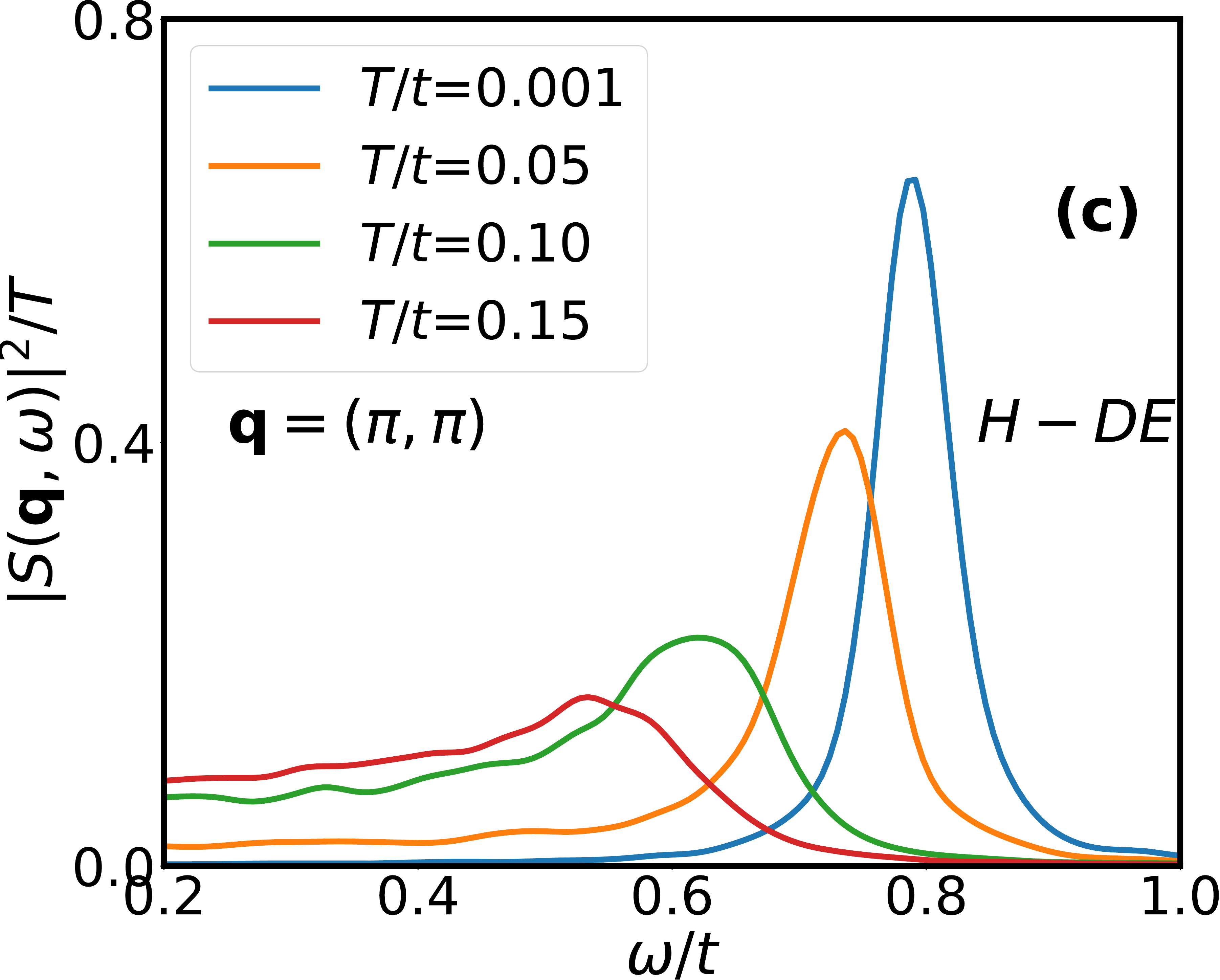}
\includegraphics[width=4.1cm,height=3.7cm]{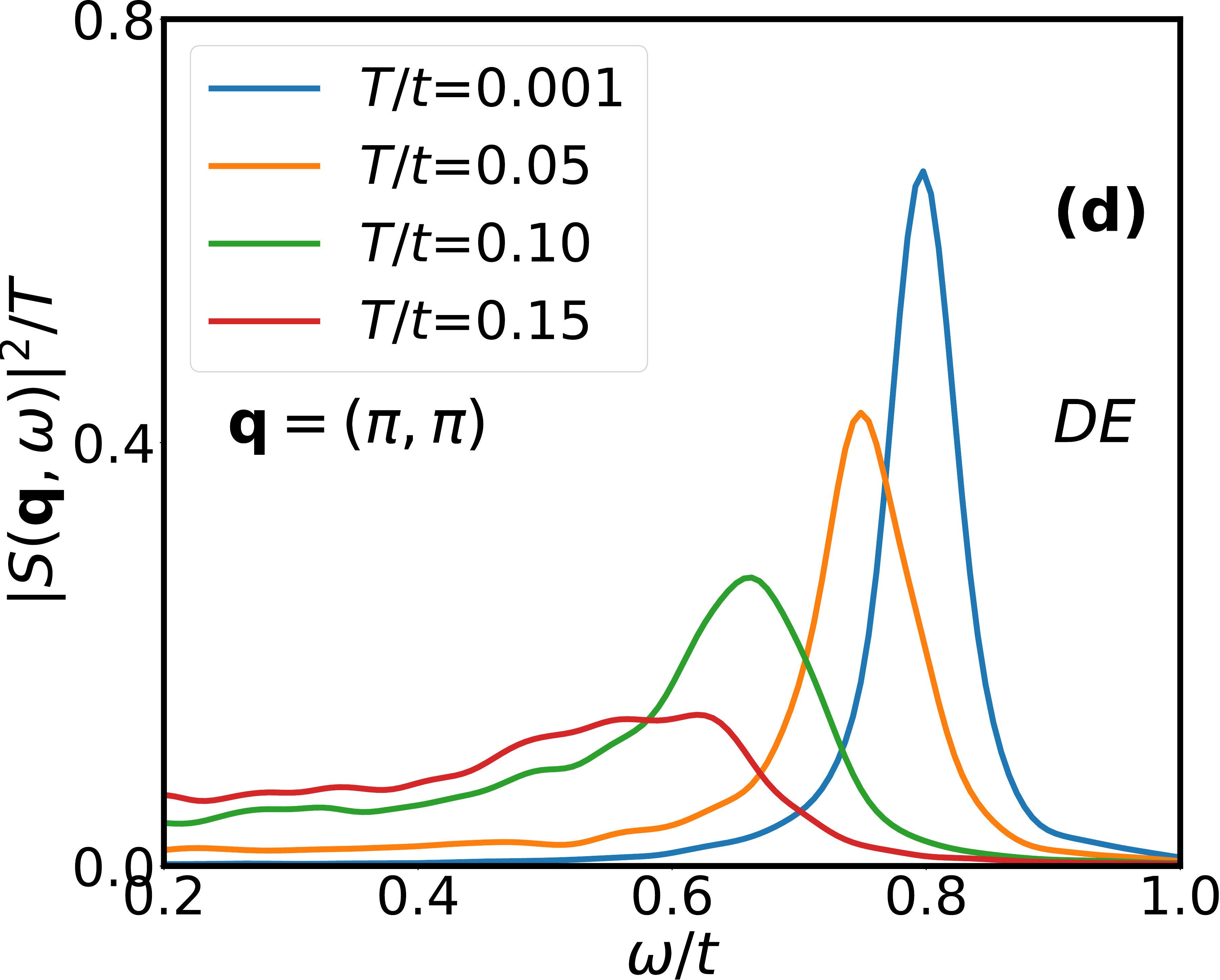}
}
\centerline{
\includegraphics[width=4.05cm,height=3.8cm]{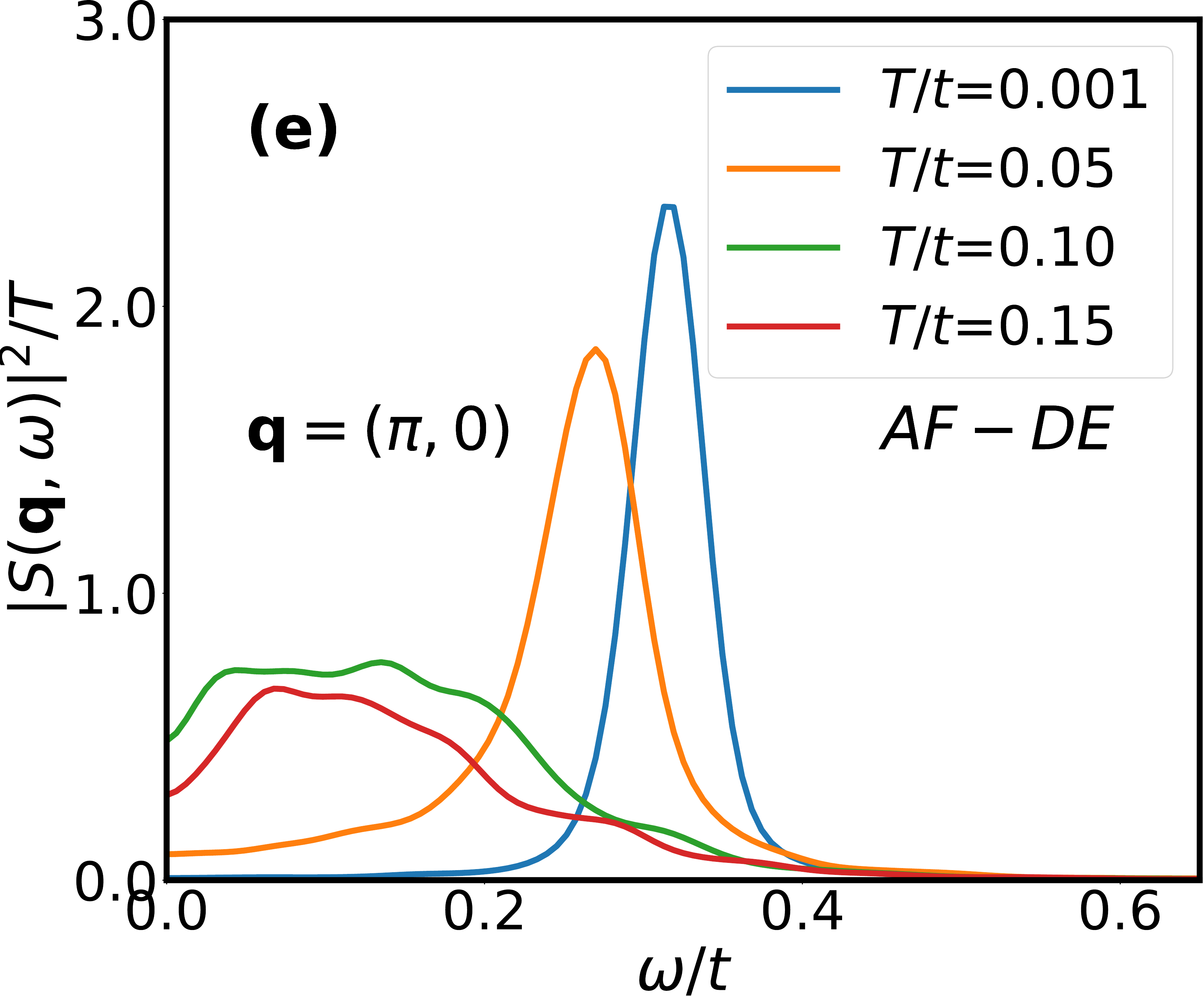}
\includegraphics[width=4.05cm,height=3.8cm]{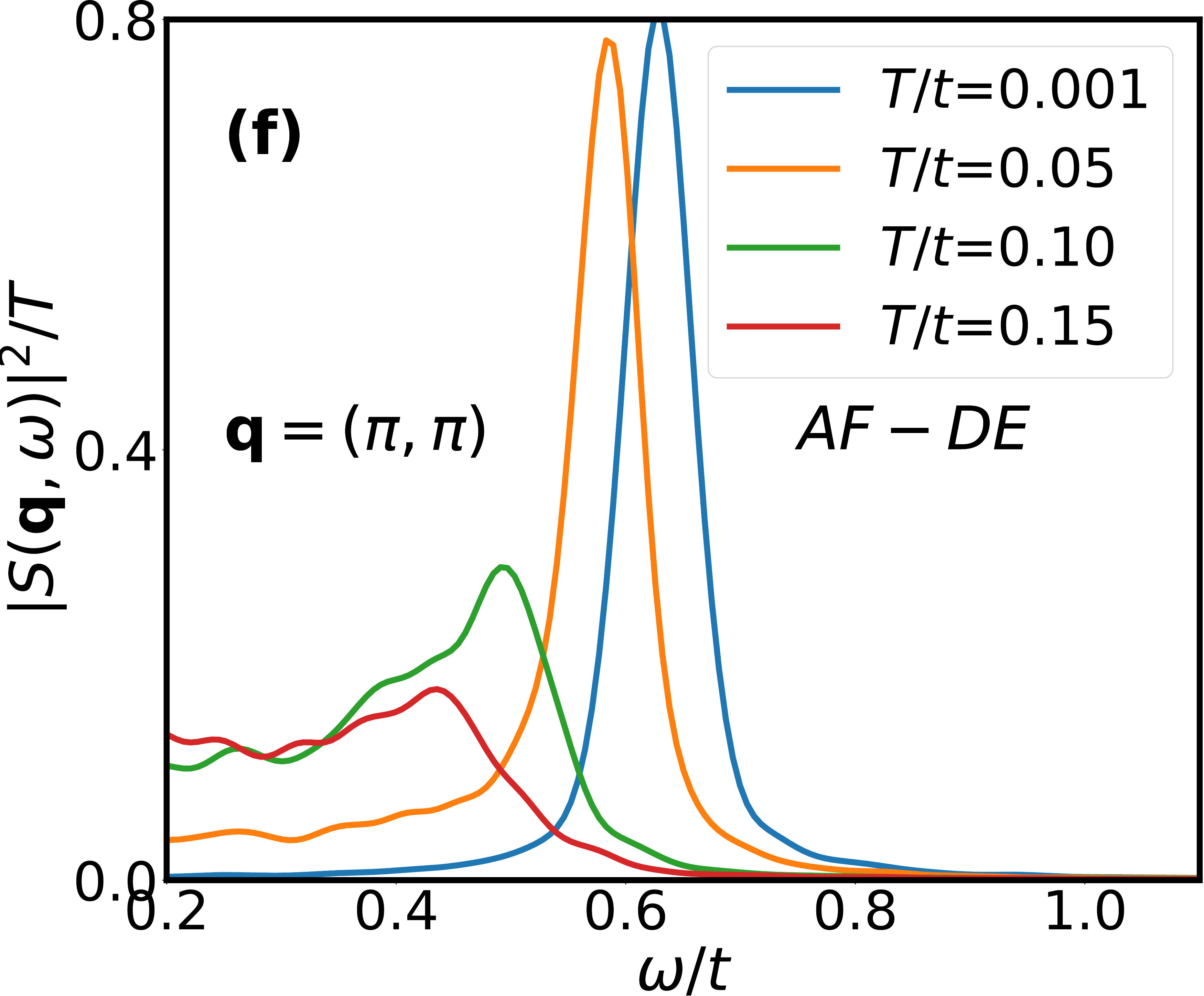}
~}
\caption{(a)-(b): Comparison of magnon lineshapes at 
${\bf q}=(\pi,0)$ for different temperature regimes 
between the present model and a pure double exchange 
model at comparable density.
(c)-(d): Same for ${\bf q}=(\pi,\pi)$.
One observes a qualitative similarity
in all regimes, signifying the subdued effect 
of magnon-phonon coupling on magnons.
(e)-(f): Magnon spectra at the same wavevectors
for a DE model with small ($J_{AF}=0.2J_{eff}$)
AF coupling. We observe a bandwidth shortening at
low $T$ and smaller dampings compared to the former.
}
\end{figure}

\section{Distribution of displacement and density}

In Fig.2, we show 
the distribution of displacements ($P(x,T)$) 
and electron density $P(n,T)$ in four temperature regimes. 
The low temperature state is uniform and
has a sharp, unimodal profile for both of them. 
The former shows gradual broadening thereafter, 
not underscoring the thermal polaron formation.
However, the density distribution 
broadens quickly ($T/T_{FM}=0.5$), 
signifying the onset of highly anharmonic behaviour. 
Near $T_{FM}$, a left-skewed, long-tailed
shape is seen, which depicts thermally induced polarons. 
Finally, beyond $T_{FM}$, a broad, plateau-like structure 
emerges, hinting at the merging of distortions of different sizes.

\begin{figure*}[t]
\centerline{
\includegraphics[width=3.9cm,height=3.6cm]{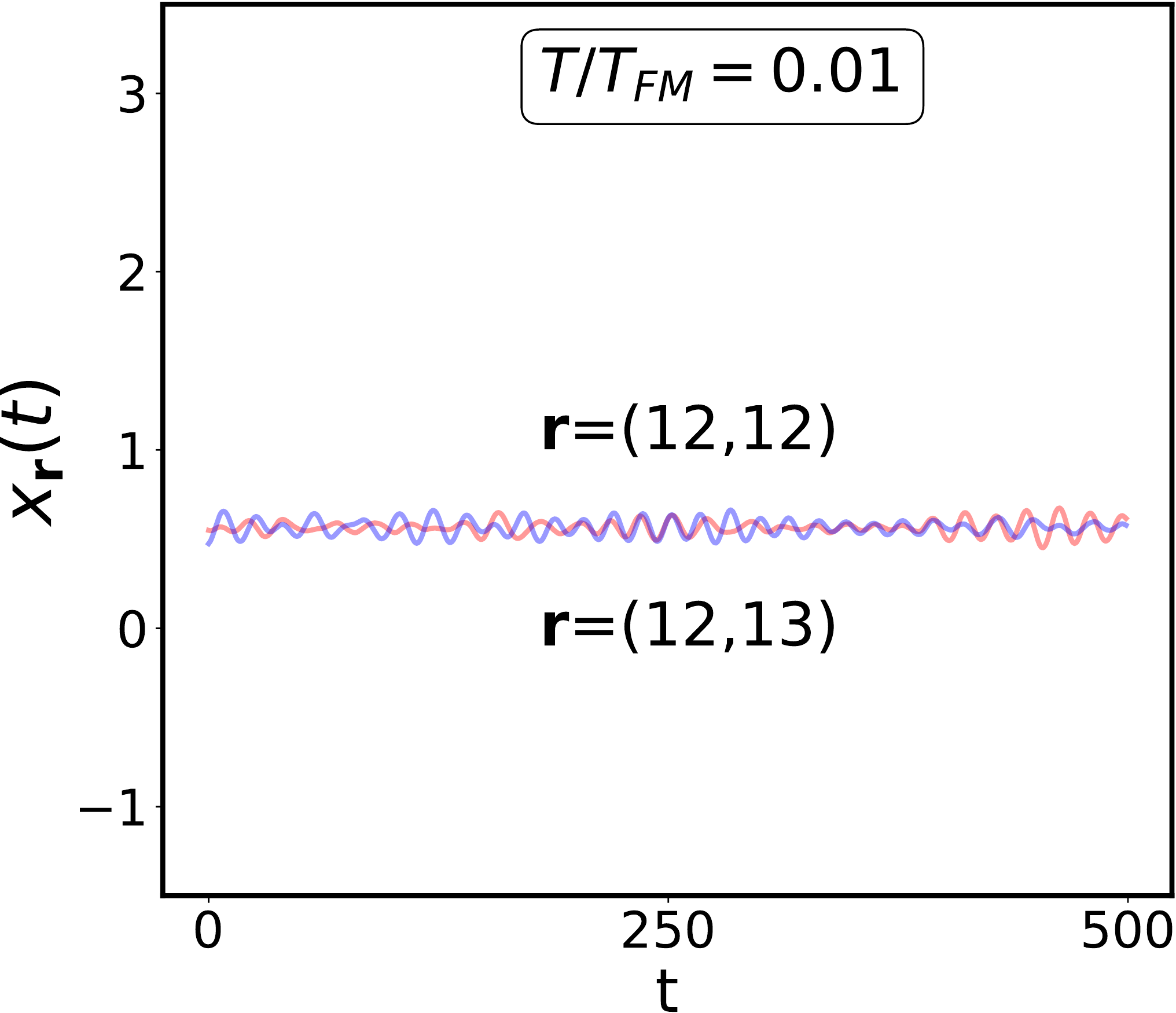}
\includegraphics[width=3.9cm,height=3.6cm]{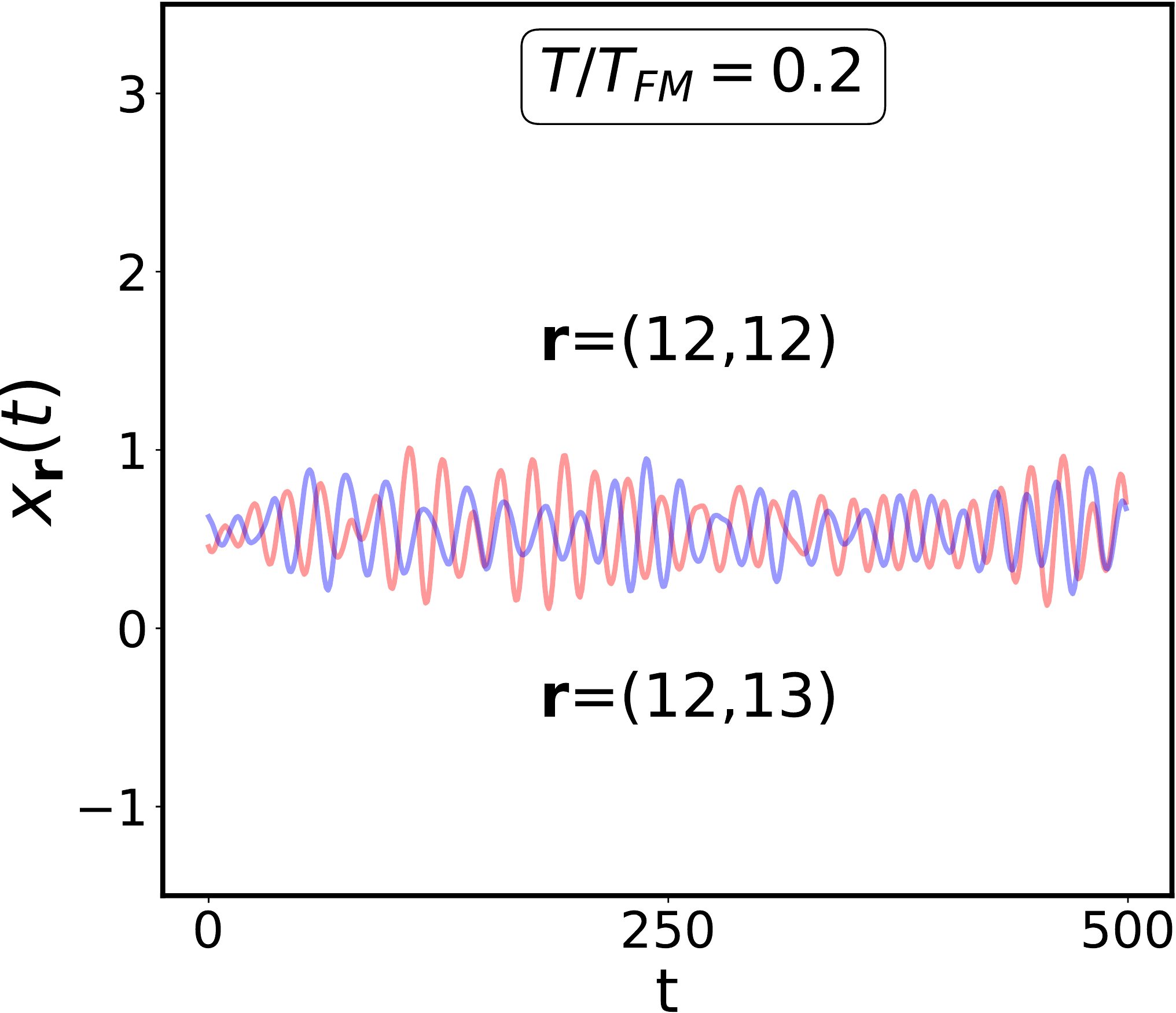}
\includegraphics[width=3.9cm,height=3.6cm]{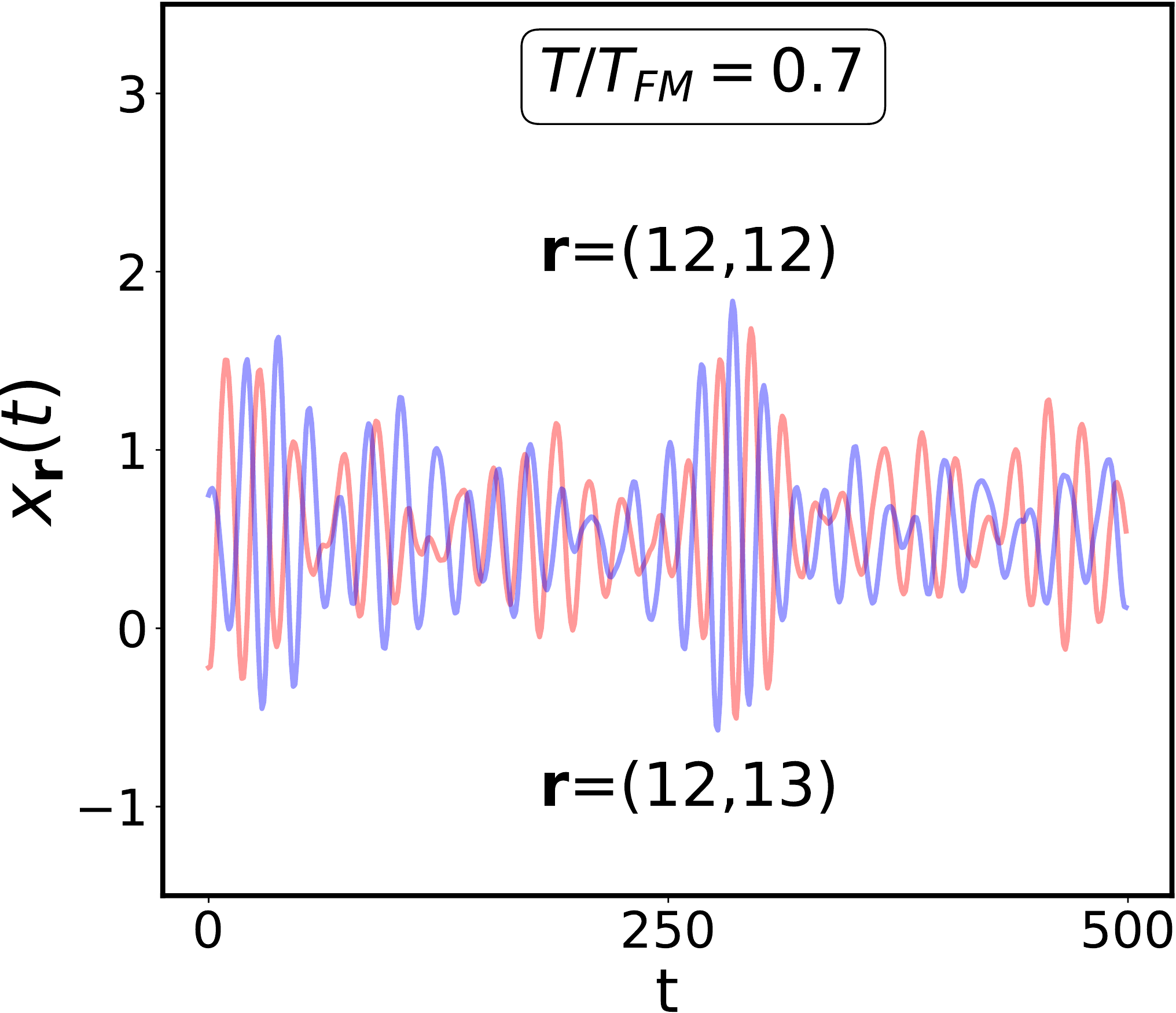}
\includegraphics[width=3.9cm,height=3.6cm]{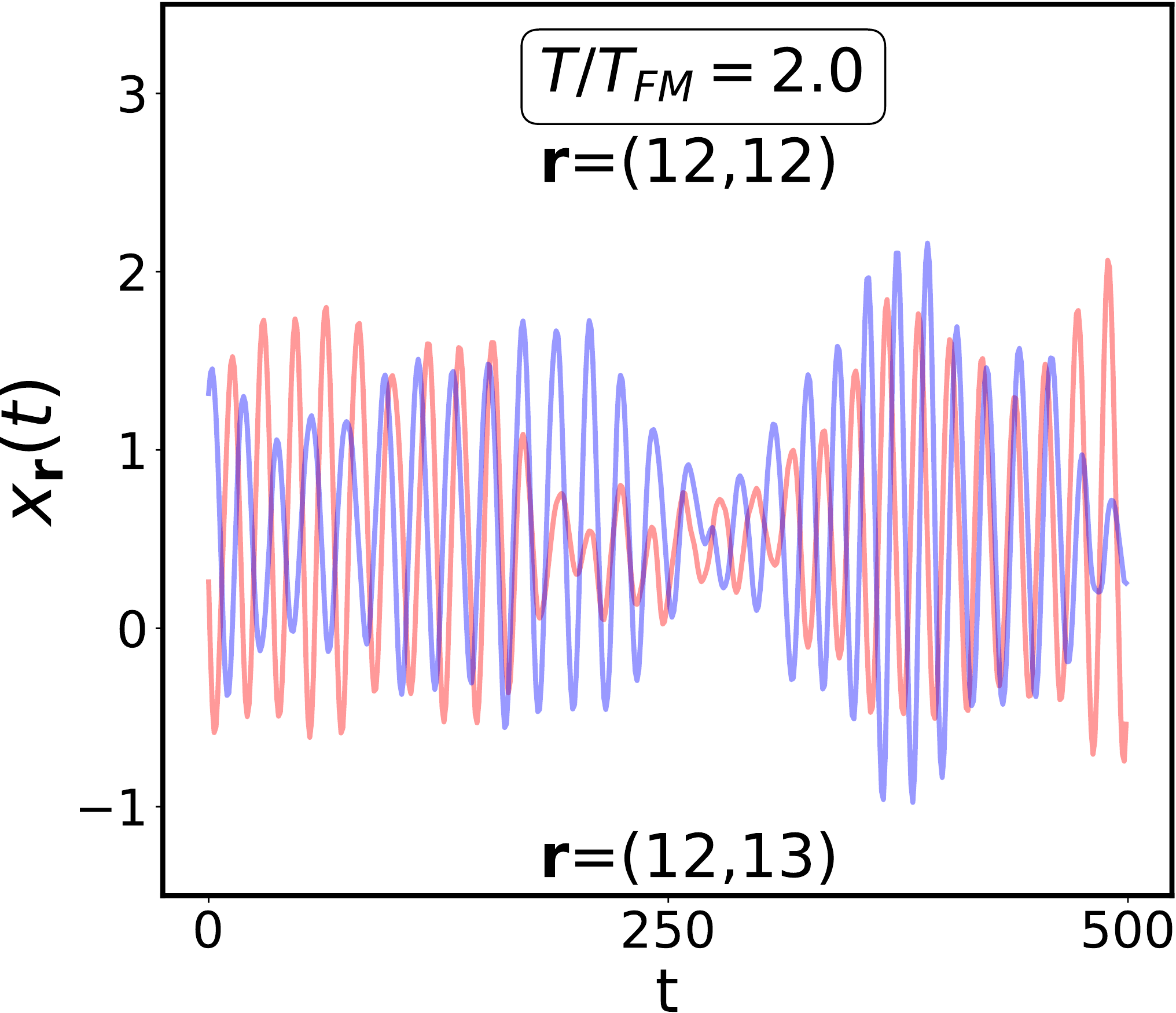}
}
\centerline{
\includegraphics[width=3.9cm,height=3.6cm]{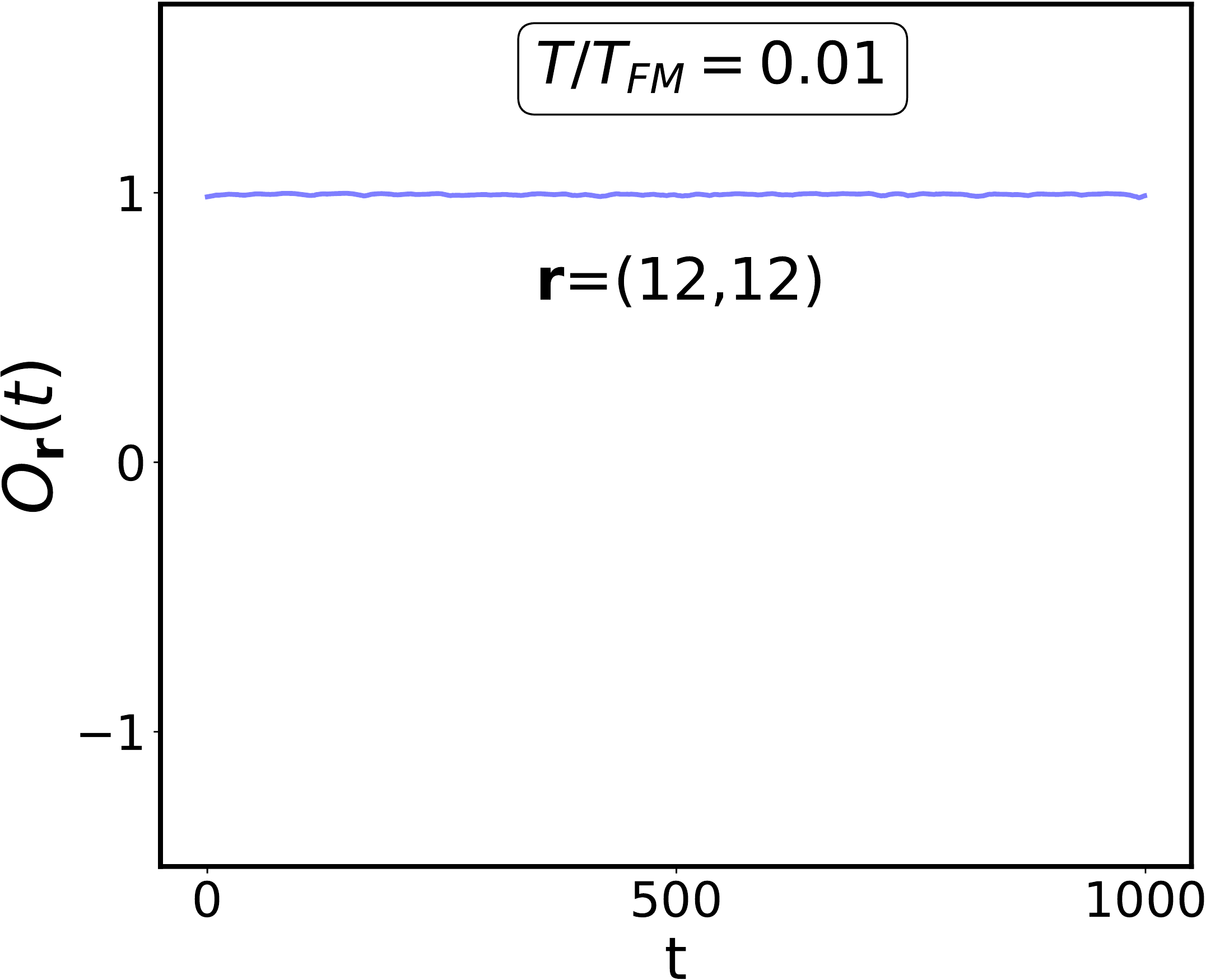}
\includegraphics[width=3.9cm,height=3.6cm]{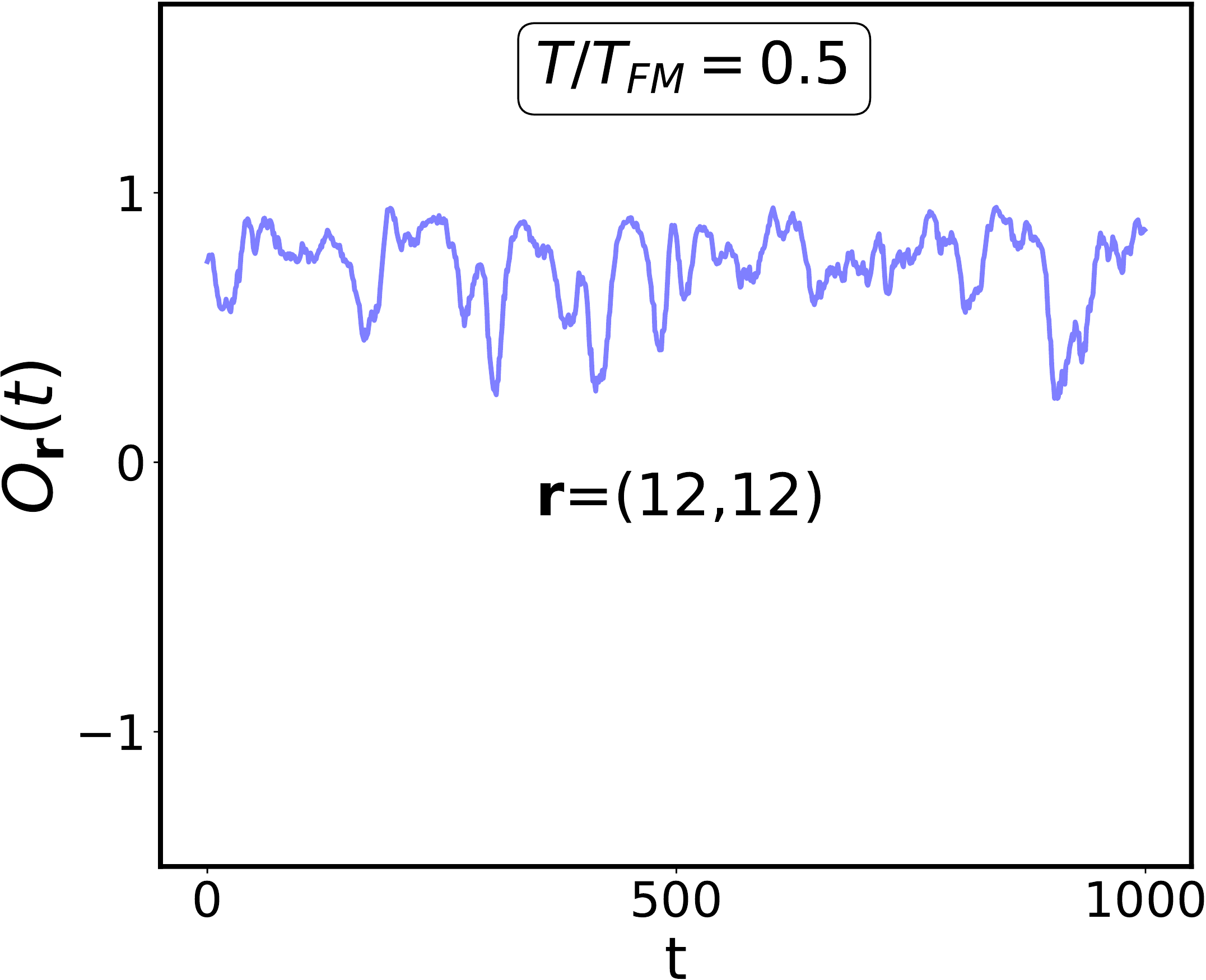}
\includegraphics[width=3.9cm,height=3.6cm]{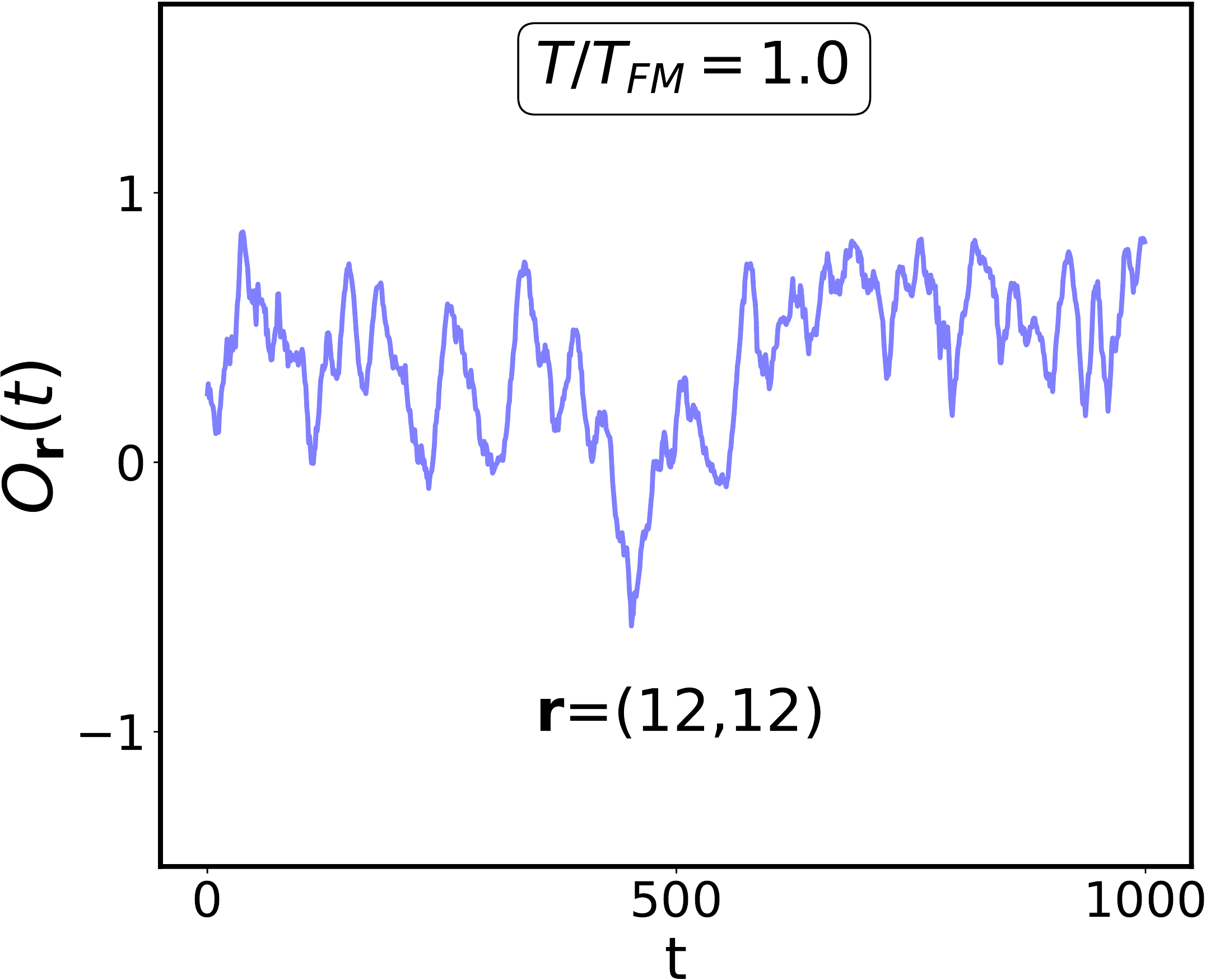}
\includegraphics[width=3.9cm,height=3.6cm]{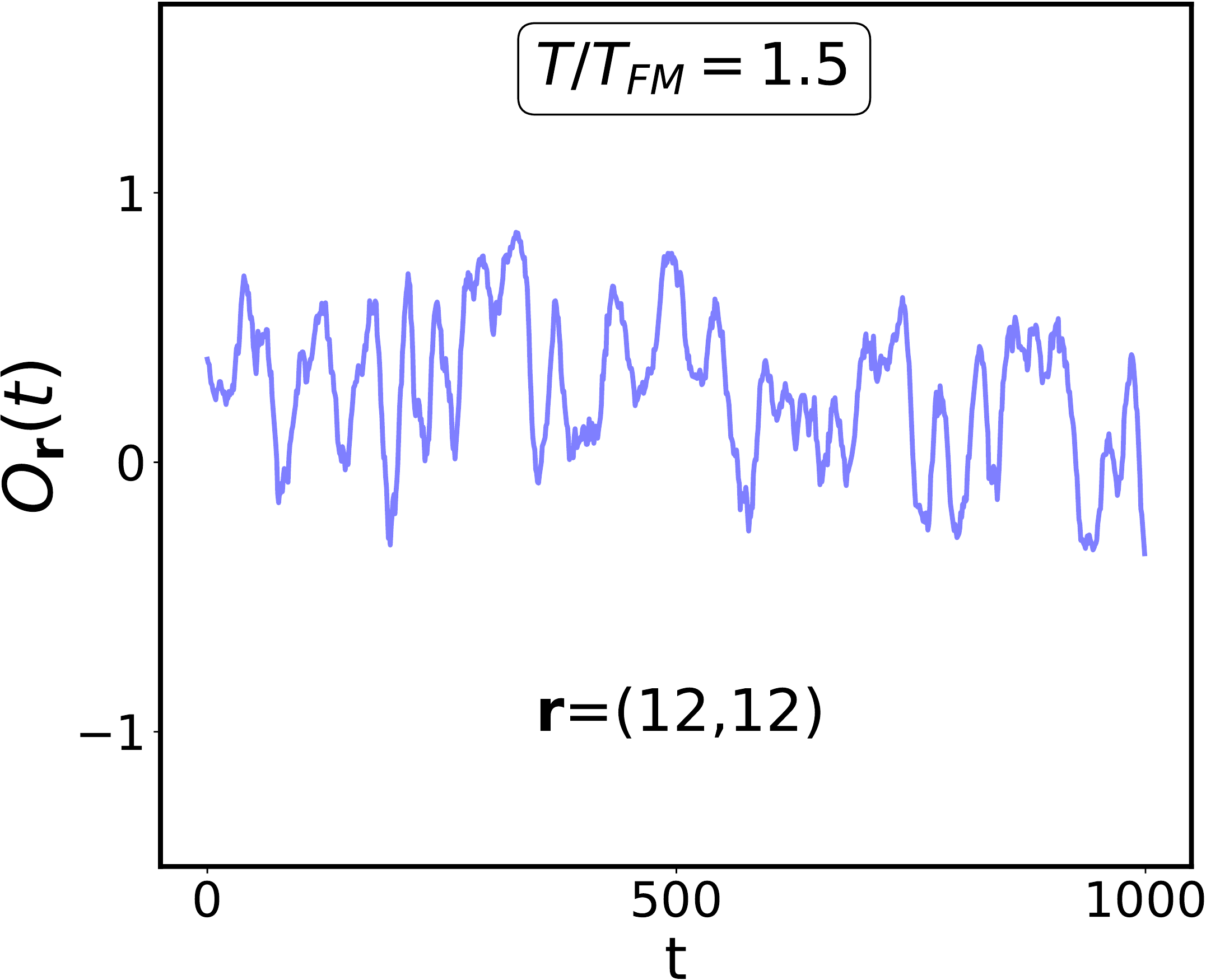}
}
\caption{Top panel: Real-time trajectories of $x_{i}(t)$ on the 
central site at four temperatures- $T/T_{FM}={0.01,0.5,1.0,1.5}$.
One observes a gradual increase in the amplitude of oscilations on
heating up. The `flip' moves, involving large ($\sim O(g/K)$) changes
in $x_{i}$, are merged with large oscillations near $T_{FM}$.
They first start showing up near $\sim0.7 T_{FM}$.
Bottom: Trajectories of the magnetic overlap function $O_{i}(t)$ on the
same site. Here, a proliferation of `sign changes' are seen with a
large period close to criticality.
}
\end{figure*}

\section{Phonon spectra in the 
Holstein-double exchange and Holstein models}

In Fig.3 we first compare lineshapes in four temperature regimes
between the Holstein-double exchange (H-DE) and the Holstein (H) model 
at comparable density. (a) and (b) feature the `high energy' spectra.
The basic trends are similar. However, the
extent of softening and broadening close to $T_{FM}$ in the former is
much more accentuated compared to the Holstein model for the same 
temperature. The basic reason is the presence of increasing spin
disorder in the former which effectively reduces the electronic 
bandwidth and enhances the effective coupling to phonons. 
However, we comment that the quantitative increase in damping
is underestimated by just considering this effect, which hints at
a crucial role of spatial correlations. The middle panel ((c) and (d))
exhibits the thermal evolution of low-frequency weight, which is
again sharper in the H-DE case, due to well-formed polarons and
their associated tunneling.
In the bottom panel ((e) and (f)), 
the softenings and linewidths extracted from the corresponding 
lineshapes are shown ((c) and (d)) . We see a more detailed thermal 
comparison here, which corroborates the above findings.
The main result is a gradual and substantial rise in 
phonon damping near $T_{FM}$ in the present model, which is absent 
in the pure Holstein case, where damping saturates by $0.5T_{FM}$. 
Moreover, the rise in softening is sharper in the pure Holstein case.

\section{Magnon spectra in the 
Holstein-double exchange and double exchange models}

Fig.4 highlights the comparison of magnon spectrum for the 
H-DE model with that of the DE model. 
The top and middle panels feature lineshapes at ${\bf q}=(\pi,0)$ and
${\bf q}=(\pi,\pi)$ respectively. The qualitative features remain
the same, which confirms that the effects of magnon-phonon coupling
on the spin dynamics are weak. Moreover, the 
double-exchange model spectra are qualitatively similar to those of a 
nearest-neighbour Heisenberg model with $J_{eff} \sim 0.1t$. This
means that the magnon dynamics is fairly conventional.
In the bottom panel ((e) and (f)), we show lineshapes at the
same wavevectors for a DE model with a small ($J_{AF}=0.2J_{eff}$)
AF coupling, which is realistic from the materials point of view.
The main result is a smaller low $T$ bandwidth and finite temperature
linewidths limited by this scale. The $T_{FM}$ in this case is 
$\sim0.08t$.

\section{Real time trajectories}

In Fig.5, we depict the representative real time trajectories 
from which we obtain the phonon and magnon power spectrum.
The top row shows $x_i(t)$ while the lower row shows the 
magnetic overlap function:
$$
O_i(t) = {1 \over 4} \sum_{\delta} {\vec S}_i(t).{\vec S}_{i + \delta}(t)
$$
where $\delta$ are the four neighbours of site $i$.
We show results in 
four temperature regimes- harmonic, anharmonic, polaronic
and high for phonons and low, intermediate, critical and
high for magnons. The chosen site is located at the middle 
of the system. The top panel also follows its right neighbour.

In the harmonic regime, the phonon dynamics 
is saturated by small oscillations about a 
homogeneous state. Anharmonic behaviour sets in
quickly ($T\sim0.2T_{FM}$), where oscillations 
of larger amplitude ($>10\%$ of mean value) 
prevail and effect of mode coupling shows up 
in $\Gamma_{ph}({\bf q})$. Next comes the 
polaronic regime ($T/T_{FM}=0.7$), 
where short-range correlated `burst' like events 
coexist with anharmonic oscillations. These
lead to the remarkable low-frequency weight
transfer. Finally, well beyond $T_{FM}$, 
distortions of several scales get merged 
together and short-range correlations weaken.

The magnetic overlap $O_{i}(t)$ is perfect at
low $T$, but shows a gradual 
proliferation of `sign changing' moves on heating,
which eventually kill the ferromagnetic order.

\bibliographystyle{unsrt}